\documentclass[12pt, draftcls]{./IEEEtran_v15}%
\usepackage{amssymb}
\usepackage{amsmath}
\usepackage{graphicx}%
\usepackage{amsfonts}
\begin{document}

\title{Interference Channel with an Out-of-Band Relay}
\author{Onur Sahin, Osvaldo Simeone and Elza Erkip\\\thanks{ Onur Sahin was with the Dept. of Electrical and
Computer Engineering, Polytechnic Institute of New York University, he is now with InterDigital, Inc. New York, USA (email: onur.sahin@interdigital.com). Osvaldo
Simeone is with the CWCSPR, New Jersey Institute of Technology, Newark, NJ
07102, USA (email:osvaldo.simeone@njit.edu). Elza Erkip is with the Dept. of Electrical and
Computer Engineering, Polytechnic Institute of New York University, Brooklyn,
NY 11201, USA (email: elza@poly.edu). This work is partially supported
by NSF grant No. 0520054, The Wireless Internet Center for Advanced Technology
at Polytechnic Institute of NYU. The work of O. Simeone is partially supported
by CCF-0914899.}}
\maketitle

\begin{abstract}
A Gaussian interference channel (IC) with a relay is considered. The relay is
assumed to operate over an orthogonal band with respect to the underlying IC,
and the overall system is referred to as IC with an out-of-band relay
(IC-OBR). The system can be seen as operating over two parallel
interference-limited channels: The first is a standard Gaussian IC and the
second is a Gaussian relay channel characterized by two sources and
destinations communicating through the relay without direct links. We refer to
the second parallel channel as OBR Channel (OBRC).

The main aim of this work is to identify conditions under which optimal
operation, in terms of the capacity region of the IC-OBR, entails either
\textit{signal relaying} and/or \textit{interference forwarding} by the relay,
with either a separable or non-separable use of the two parallel channels, IC
and OBRC. Here \textquotedblleft separable\textquotedblright\ refers to
transmission of independent information over the two constituent channels.

For a basic model in which the OBRC consists of four orthogonal channels from
sources to relay and from relay to destinations (IC-OBR\ Type-I), a condition
is identified under which signal relaying and separable operation is optimal.
This condition entails the presence of a \textit{relay-to-destinations
capacity bottleneck} on the OBRC and holds irrespective of the IC. When this
condition is not satisfied, various scenarios, which depend on the IC\ channel
gains, are identified in which interference forwarding and non-separable
operation are necessary to achieve optimal performance. In these scenarios,
the system exploits the ``excess capacity'' on the OBRC via interference
forwarding to drive the IC-OBR system in specific interference regimes (strong
or mixed).

The analysis is then turned to a more complex IC-OBR, in which the OBRC
consists of only two orthogonal channels, one from sources to relay and one
from relay to destinations (IC-OBR Type-II). For this channel, some capacity
resuls are derived that parallel the conclusions for IC-OBR\ Type-I and point
to the additional analytical challenges.

\end{abstract}

%EndAName
%

%TCIMACRO{\TeXButton{Def}{\def\path{C:/ARTICOLI/Radiomobile/papers/JST/Figure/}%
%}}%
%BeginExpansion
\def\path{C:/ARTICOLI/Radiomobile/papers/JST/Figure/}%
%EndExpansion

%EndAName%
%TCIMACRO{\TeXButton{Sloppy}{\sloppy}}%
%BeginExpansion
\sloppy
%EndExpansion

\section{Introduction\label{sec_introduction}}

Modern wireless communication networks are characterized by the coexistence of
an increasing number of interfering devices and systems. While this often
leads to an overall system performance that is limited by mutual interference,
the presence of many independent wireless devices may also potentially offer
new opportunities and performance benefits by allowing \textit{cooperation}.
Opportunities for cooperation are further enhanced for \textit{multistandard}
terminals that are able to communicate simultaneously over multiple radio
interfaces, and thus to interact and cooperate with devices belonging to
different systems and networks. For instance, many current wireless terminals
are equipped with a 3G cellular transceiver along with a Wi-Fi interface. This
paper focuses on investigating the advantages of cooperation in
interference-limited scenarios where cooperation is enabled by orthogonal
radio interfaces and multistandard terminals.

\begin{figure}[t]
\begin{center}
\includegraphics[width=10 cm]{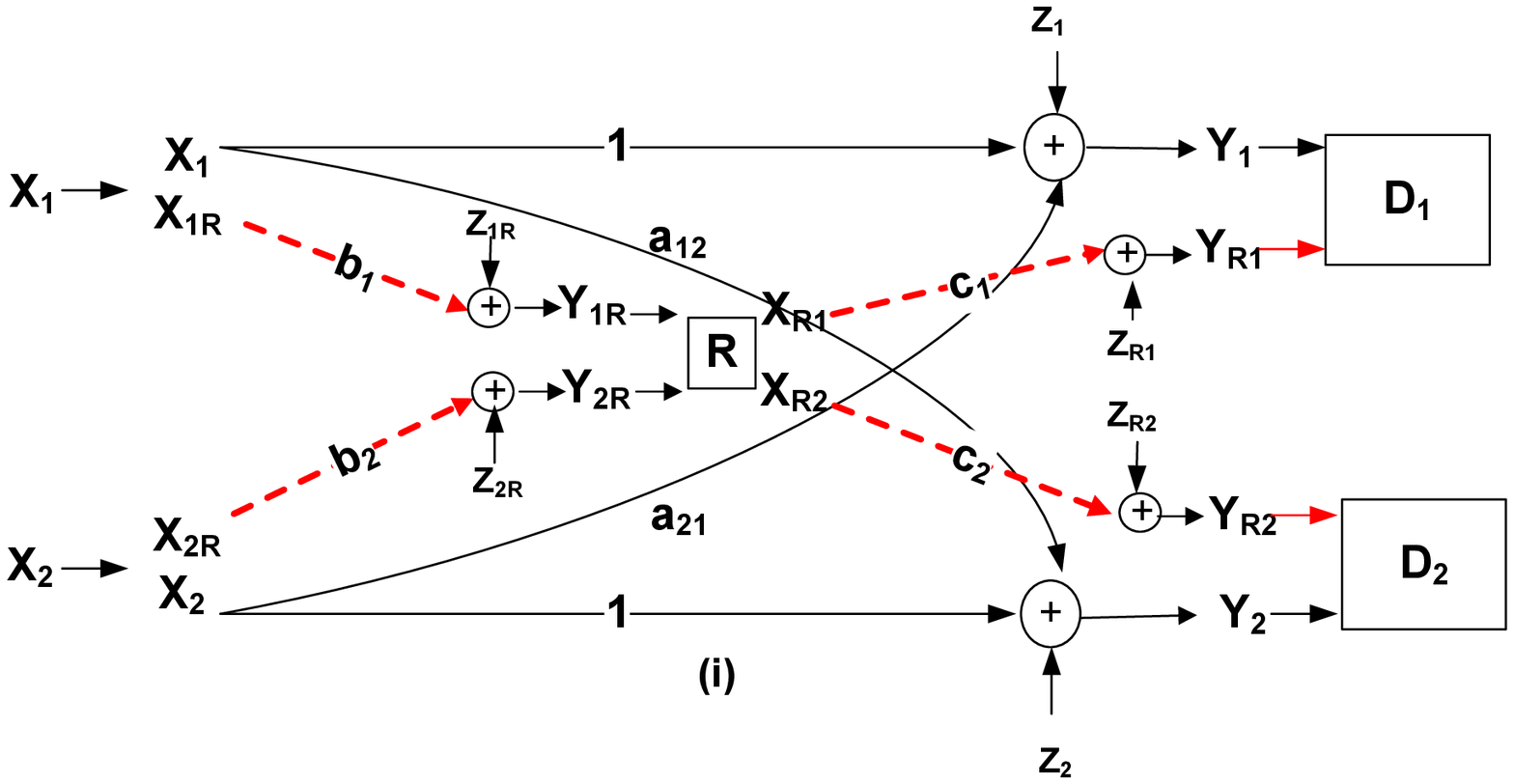} \includegraphics[width=10
cm]{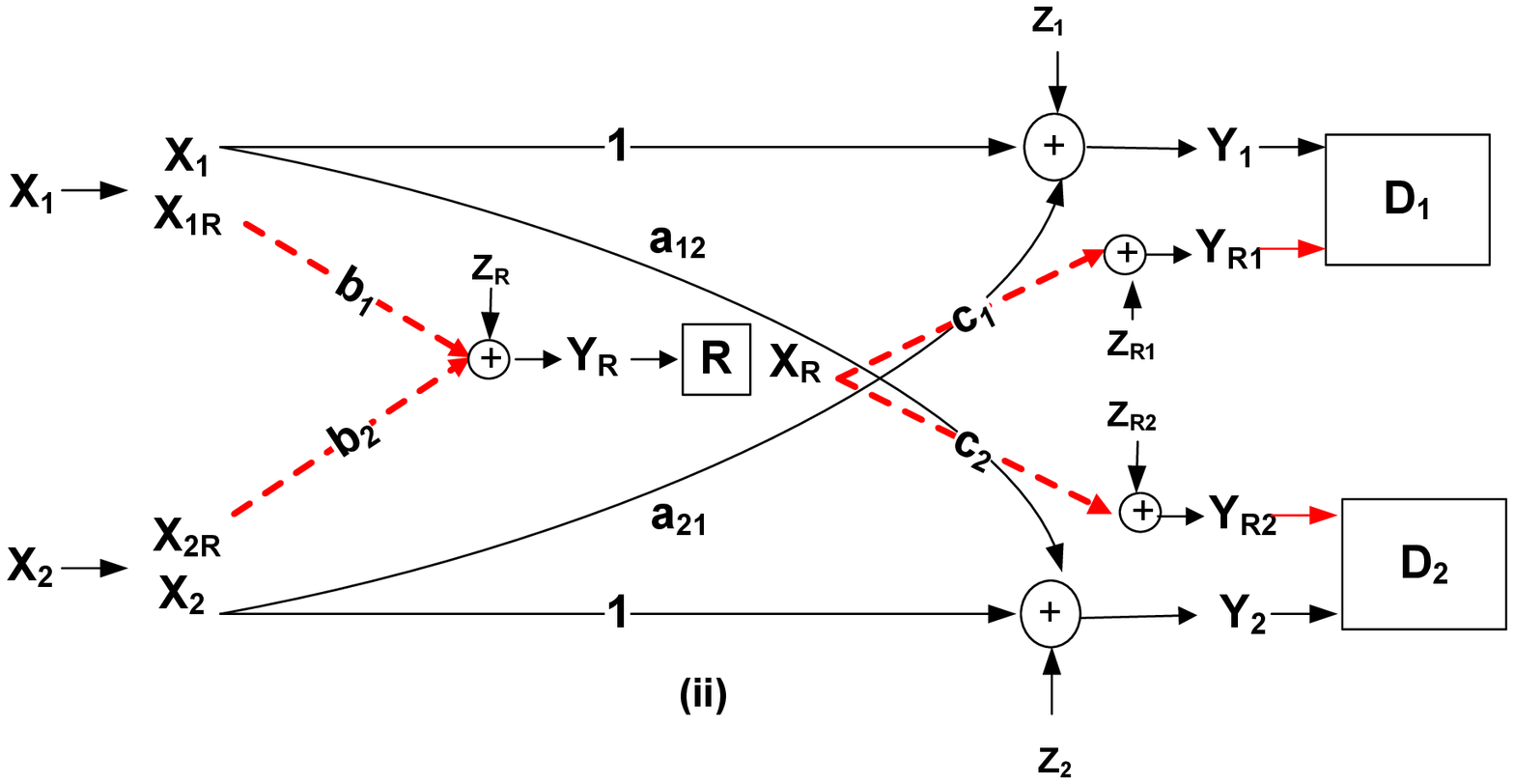}
\end{center}
\caption{Interference Channel (IC) with an out-of-band relay (OBR). The OBR
channel (OBRC) has $\eta$ channel uses for each channel use of IC.
(\textit{i}) IC-OBR Type-I: The OBRC is divided into four Gaussian\ orthogonal
channels with $\eta_{iR}$, $\eta_{Ri}$, $i=1,2$, channel uses each;
(\textit{ii}) IC-OBR Type II: The OBRC is divided into two orthogonal channels
with $\eta_{MAC}$, $\eta_{BC}$ channel uses.}%
\label{Fig:System}%
\end{figure}

To fix the ideas on the problem of interest, consider Fig. \ref{Fig:System},
where a two-user Interference Channel (IC) operating over a certain bandwidth
and using a certain radio communication standard (say, Wi-Fi) is aided by a
relay that operates over an orthogonal bandwidth and uses a possibly different
standard (say, Bluetooth). For this reason, we refer to the relay as an
Out-of-Band Relay (OBR). Sources and destinations are assumed to be
multistandard, thus being able to transmit and receive, respectively, over
both radio interfaces. The system can be seen as being characterized by two
parallel channels, in which the first (i.e., the Wi-Fi channel) is a standard
Gaussian IC, and the second (i.e., the Bluetooth channel) is a particular
Gaussian relay channel. Specifically, the second channel is characterized by
two sources and destinations communicating through the relay without the
direct links between sources and destinations \cite{Nosratinia}-\cite{El Gamal
ISIT09}. We refer to it as OBR Channel (OBRC)\footnote{Notice that, while
having been studied by itself in some previous works \cite{Nosratinia}%
-\cite{El Gamal ISIT09}, as discussed below, the relay channel that forms the
OBRC does not have an agreed upon name (e.g., \cite{Nosratinia} refers to it
as "gateway" channel and \cite{El Gamal ISIT09} as\ "switch" channel).}. The
absence of a direct link on the OBRC\ can be justified in case the wireless
radio interface used on the OBRC\ has a shorter range than the one used over
the IC, as it is the case for the Wi-Fi/ Blueetooth example. We are interested
in the optimal operation on the overall system, IC and OBRC, which is termed
IC with an OBR or IC-OBR.

As explained above, the two components of an IC-OBR model are an IC (see, e.g,
\cite{Raul} for a summary of the state of the art on the subject) and the
OBRC. Assuming a half-duplex relay, the latter is modelled in two different
ways (see Fig.\ref{Fig:System}): (\textit{i}) IC-OBR Type-I:\ The OBRC is
operated by assuming orthogonal transmissions (e.g., via TDMA\ or FDMA) over
the four Gaussian links connecting sources to relay and relay to destinations;
(\textit{ii}) IC-OBR\ Type-II: The OBRC\ is more generally operated by
orthogonalizing the Gaussian multiple access channel between the two sources
and the relay and the broadcast channel from relay to destinations. It is
noted that relay models similar to the assumed OBRC have been recently studied
in \cite{Nosratinia} from an outage perspective and in \cite{Guduz
ISIT09}\cite{El Gamal ISIT09} assuming (symmetric) two-way communications (see
also \cite{wilson_lattice}\cite{nam_lattice}). Overall, it is remarked that
general capacity results are unavailable for both component channels of a
IC-OBR, namely the IC and OBRC.

Beside the connection with the literature on the two individual component
channels, briefly reviewed above, the considered model is related to, and
inspired by, two recent lines of work. The first deals with \textit{relaying
in interference-limited systems}, where, unlike the IC-OBR, the relay is
assumed to operate in the same band as the IC \cite{onur_globecom}%
-\cite{Sriram}. These works reveal the fact that relaying in
interference-limited systems offers performance benefits not only due to
\emph{signal relaying,} as for standard relay channels (see, e.g.,
\cite{cover_relay}), but also thanks to the novel idea of \emph{interference
forwarding}. According to the latter, the relay helps by either reinforcing
the interference received at the undesired destination so as to facilitate
interference stripping, or by conversely reducing the interference via
negative beamforming. Achievable schemes incorporating these operations for a
Gaussian system are developed in \cite{onur_asil07} \cite{onur_globecom}
\cite{Sriram}, while \cite{ivana_ISIT08} deals with a discrete-memoryless
model and \cite{Aylin Globecom09} derives an upper bound on the capacity
region by allocating infinite power to the relay (see also \cite{onur_ISIT09}
for a review).

A second related line of work deals with communications over \textit{parallel
ICs }(albeit the considered OBRC\ is not a conventional IC). As shown in
\cite{cadambe} \cite{Lalitha} \cite{Koreans}, optimal operation over parallel
ICs, unlike scenarios with a single source or destination, typically entails
joint coding over the parallel channels. In other words, the signals sent over
the parallel ICs need to be generally correlated to achieve optimality, and
thus a \textit{separable} approach, whereby the parallel channels are treated
independently, is in general not sufficient. To elaborate, recall that the
most general transmission scheme for a regular IC involves splitting of each
source's message into both private and common submessages, where the first is
treated as noise at the unintended destination, while the latter is jointly
decoded with the messages of the intended source. Now, in the presence of
parallel ICs, given the above, the question arises as to what type of
information, either private or common, should be sent over the parallel
channels. For instance, the original work \cite{cadambe} derives conditions
under which correlated transmission of private messages is optimal and
\cite{Lalitha} considers the optimality of common information transmission,
whereas in \cite{Koreans} scenarios are found for which sending both
correlated private and common messages is optimal. We refer to \cite{Lalitha},
\cite{Shang_noisy_parallel} and \cite{Koreans} for a summary of special cases
in which a separable approach is instead optimal.

For the IC-OBR\ under study, the main issue we are interested in is assessing
under which conditions the OBRC\ should be optimally used for either or both
\textit{signal relaying} or \textit{interference forwarding}, and by following
a\textit{\ separable }or \textit{non-separable} transmission strategy over the
parallel channels given by the IC and the OBRC. The main contribution of the
paper is the identification of a number of such scenarios, which we classify
as being characterized by \textit{relay-to-destinations bottleneck} or
\textit{excess rate conditions}, for both IC-OBR Type-I and Type-II and
assuming both fixed and variable bandwidth allocations on the OBRC.

The paper is organized as follows. In Section \ref{Sec:System Models}, we give
the system models of IC-OBR Type-I and Type-II channels. Section
\ref{Sec:ICOBR_ind_chn} gives a general outer bound and achievable region for
IC-OBR Type-I for both fixed and variable OBRC\ bandwidth allocations.
Capacity results are established for both cases. In Section
\ref{Sec:IC-OBR Type II}, we give outer bounds on the capacity region for
IC-OBR\ Type-II. Conditions for the optimality of signal relaying and/or
interference forwarding for IC-OBR\ Type-II are established in this section.
Finally, we conclude the paper in Section \ref{Sec:Conclusion}.

%\begin{figure}[t]
%\begin{center}
%\includegraphics[width=10 cm]{sys_mod}
%\label{Fig:Systemb}
%\end{center}
%\caption{System Model}
%\end{figure}

\textit{Notation}: We define $\mathcal{C}(x)=1/2\log_{2}(1+x)$.

\section{System Models}

\label{Sec:System Models} We investigate the IC-OBR models shown in Fig.
\ref{Fig:System}, which we refer to as IC-OBR Type-I and Type-II. In both
models, the sources $S_{1}$ and $S_{2}$ communicate to their respective
destinations $D_{1}$ and $D_{2}$ via two orthogonal channels, namely a
Gaussian IC and the out-of-band relay channel (OBRC), where the latter is
characterized by $\eta\geq0$ channel uses per channel use of the IC. In
pratice, parameter $\eta$ can be thought of as the ratio between the bandwidth
of the OBRC\ and of the IC. Specifically, each source $S_{i},$ $i=1,2,$ wishes
to send a message index $W_{i},$ uniformly drawn from the message set
$[1,2^{nR_{i}}]$ \footnote{As\ it is common in the literature, we consider the
number of messages $2^{nR_{i}}$ rounded off to the smallest larger integer. We
will use the same convention wherever integer quantities are needed.}$,$ to
its destination $D_{i},$ with the help of an OBR, which operates half-duplex.
Notice that $n$ is the number of channel uses of the IC available for
communication of the given messages (which yields $\eta n$ channel uses for
the OBRC), so that $R_{i}$ is the rate of the $i$th pair ($S_{i},D_{i}$) in
terms of bits per IC channel use.

The signals received on the IC by the two receivers $D_{1}$ and $D_{2}$ in
channel use $t=1,...,n$ are given as, respectively,
\begin{subequations}
\label{sys_mod_a}%
\begin{align}
Y_{1,t} &  =X_{1,t}+a_{21}X_{2,t}+Z_{1,t}\\
Y_{2,t} &  =a_{12}X_{1,t}+X_{2,t}+Z_{2,t},
\end{align}
where $X_{i,t}\in\mathbb{R}$ represents the (real) input symbol of source
$S_{i}$, which satisfies the power constraint $\frac{1}{n}\sum_{t=1}%
^{n}E[X_{i,t}^{2}]\leq P_{i}$, and $Z_{i,t}$ are independent identically
distributed (i.i.d.) zero-mean Gaussian noise processes with unit power. The
two IC-OBR\ models studied in the following differ in the way the OBRC is
operated, as discussed below. Both models assume a half-duplex relay.

\subsection{IC-OBR Type-I}

In the IC-OBR Type-I model, shown in Fig.\ref{Fig:System}-(\textit{i}), the
OBRC bandwidth (or equivalently the set of channel uses $\eta n$) is
partitioned into four orthogonal Gaussian channels, corresponding to different
source-to-relay and relay-to-destination pairs. This can be realized by
orthogonal access schemes such as TDMA\ or FDMA. Specifically, we have two
Gaussian channels from sources $S_{i}$ to relay $R$ with fraction of channel
uses $\eta_{iR}$, $i=1,2,$ and two Gaussian channels from the relay $R$ to
destinations $D_{i},$ $i=1,2,$ with fraction of channel uses $\eta_{Ri}.$ We
have $\sum_{i=1}^{2}(\eta_{iR}+\eta_{Ri})=\eta$. The signals received by the
relay $R$ over the OBRC on the source-to-relay channels are given by
\end{subequations}
\begin{equation}
Y_{iR,t}=b_{i}X_{iR,t}+Z_{iR,t},\label{YiR}%
\end{equation}
for $t=1,...,\eta_{iR}n$ and $i=1,2,$ whereas the signals received at the
destination over the OBRC on the relay-to-destination channels are given by
\begin{equation}
Y_{Ri,t}=c_{i}X_{Ri,t}+Z_{Ri,t},
\end{equation}
for $t=1,...,\eta_{Ri}n$, $i=1,2,$ where ($Z_{iR,t},Z_{Ri,t})$ are i.i.d.
zero-mean Gaussian noise processes with unit power. We assume power
constraints $\frac{1}{n}\sum_{t=1}^{\eta_{iR}n}E[X_{iR,t}^{2}]\leq P_{iR}$,
and $\frac{1}{n}\sum_{t=1}^{\eta_{Ri}n}E[X_{Ri,t}^{2}]\leq P_{Ri}$, $i=1,2$.
The rationale behind this power constraint assumption arises for the scenarios
such as the relay transmits using TDMA with per-symbol power constraints, or
employs FDMA transmission with spectral mask constraints. Another model that
encompasses this assumption is where the relay communicates with the
destinations using two distinct radio interfaces with different transceivers.

A $(2^{nR_{1}},2^{nR_{2}},n)$ code for the IC-OBR type-I is defined by:
(\textit{a}) The encoding functions at the sources $S_{i},$ $i=1,2,$ given by
\begin{equation}
f_{i}^{(n)}\text{: }[1,2^{nR_{i}}]\rightarrow\mathbb{R}^{n}\times
\mathbb{R}^{\eta_{iR}n}%
\end{equation}
which maps a message $W_{i}\in\lbrack1,2^{nR_{i}}]$ into the codewords
($X_{i}^{n},X_{iR}^{\eta_{iR}n})=f_{i}^{(n)}(W_{i})$ to be transmitted on the
IC and OBRC, respectively; (\textit{b}) The encoding function $g^{(n)}$ at the
relay given by
\begin{equation}
g^{(n)}\text{: }\mathbb{R}^{\eta_{1R}n}\times\mathbb{R}^{\eta_{2R}%
n}\rightarrow\mathbb{R}^{\eta_{R1}n}\times\mathbb{R}^{\eta_{R2}n},
\end{equation}
which maps the received signal to the codewords ($X_{R1}^{n\eta_{R1}}%
,X_{R2}^{n\eta_{R2}})=g^{(n)}(Y_{R1}^{\eta_{1R}n},Y_{R2}^{\eta_{2R}n})$,
$i=1,2,$ that are sent to the destinations; (\textit{c}) The decoding
functions at the destinations $D_{i}$, denoted by $h_{i}^{(n)}$, $i=1,2,$
with,
\begin{equation}
h_{i}^{(n)}\text{: }\mathbb{R}^{n}\times\mathbb{R}^{\eta_{Ri}n}\rightarrow
\lbrack1,2^{nR_{i}}]
\end{equation}
which maps the received signal via the IC, $Y_{i}^{n}$ and OBRC $Y_{Ri}%
^{\eta_{Ri}n}$ into the estimated message $\widehat{W}_{i}=h_{i}^{(n)}%
(Y_{i}^{n},Y_{Ri}^{\eta_{Ri}n})$.

\subsection{IC-OBR Type-II}

\label{Sec:Type-II}

The IC-OBR Type-II model, shown in Fig.\ref{Fig:System}-(\textit{ii}), the
OBRC\ is orthogonalized into two channels, one being a multiple-access channel
(MAC) from $S_{1}$ and $S_{2}$ to $R,$ with fraction of channel uses
$\eta_{MAC},$ and the other being a broadcast channel (BC) from $R$ to $D_{1}$
and $D_{2}$, with fraction of channel uses $\eta_{BC}$. We have $\eta
_{MAC}+\eta_{BC}=\eta$. The received signal at the relay $R$ over the OBRC\ is
given by
\begin{equation}
Y_{R,t}=b_{1}X_{1R,t}+b_{2}X_{2R,t}+Z_{R,t}%
\end{equation}
for $t=1,...,\eta_{MAC}n$; and the signal received at destination $D_{i}$ over
the OBRC is
\[
Y_{Ri,t}=c_{i}X_{R,t}+Z_{Ri,t},\mbox{  }i=1,2,
\]
for $t=1,...,\eta_{BC}n$, $i=1,2.$ We have the power constraints $\frac{1}%
{n}\sum_{t=1}^{\eta_{MAC}n}E[X_{iR,t}^{2}]\leq P_{iR}$, and $\frac{1}{n}%
\sum_{t=1}^{\eta_{BC}n}E[X_{R,t}^{2}]\leq P_{R}$.

A $(2^{nR_{1}},2^{nR_{2}},n)$ code for the IC-OBR Type-II is defined similar
to codes for IC-OBR\ Type-I with the difference that the encoding function
$g^{(n)}$ at the relay is modified as
\[
g^{(n)}\text{: }\mathbb{R}^{\eta_{MAC}n}\rightarrow\mathbb{R}^{\eta_{BC}n},
\]
which maps the received signal $Y_{R}^{\eta_{MAC}n}$ into the transmitted
codeword $X_{R,t}^{\eta_{BC}n}=g^{(n)}(Y_{R}^{\eta_{MAC}n}).$

Note that IC-OBR Type-I is a special case of Type-II, obtained by
orthogonalizing the MAC and the BC.

%\subsection{Symmetric IC-OBR Type-I and Type-II}
%
%Throughout the paper an IC-OBR with symmetric IC is defined by
%\begin{equation}
%a\triangleq a_{12}=a_{21},\text{ }P\triangleq P_{1}=P_{2},\label{symmetric 1}%
%\end{equation}
%whereas for both Type-I and Type-II we will refer to the OBRC as symmetric if
%it satisfies
%\begin{equation}
%b\triangleq b_{1}=b_{2},\text{ }c\triangleq c_{1}=c_{2},\text{ }%
%P_{s}\triangleq P_{1R}=P_{2R},\text{ }P_{R}\triangleq P_{R1}=P_{R2}.
%\end{equation}

\subsection{Achievable Rates for Fixed and Variable OBRC Bandwidth Allocation}

Following conventional definitions, we define the probability of error as the
probability that any of the two transmitted messages is not correctly decoded
at the intended destination. Achievable rates ($R_{1},R_{2}$) are then defined
for two different scenarios: (\textit{a}) \textit{Fixed OBRC bandwidth
allocation}: Here, the bandwidth allocation parameters over the OBRC, namely
($\eta_{R1},\eta_{R2},\eta_{1R},\eta_{2R}$) for IC-OBR\ Type-I and
($\eta_{MAC},\eta_{BC}$) for Type-II, are considered to be given and fixed.
Therefore, a rate pair ($R_{1},R_{2}$) is achievable if a coding scheme can be
found that drives the probability of error to zero for the given
(feasible\footnote{Bandwidth allocation parameters are feasible if $\sum
_{i=1}^{2}(\eta_{Ri}+\eta_{iR})=\eta$ for IC-OBR\ Type-I and $\eta_{MAC}%
+\eta_{BC}=\eta$ for Type-II.}) bandwidth allocation parameters; (\textit{b})
\textit{Variable OBRC bandwidth allocation}: Here the bandwidth allocation can
be optimized, so that a rate pair ($R_{1},R_{2}$) is said to be achievable if
a coding scheme exists that drives the probability of error to zero for some
feasible bandwidth allocation parameters. The capacity region is in both cases
the closure of the set of all achievable rates.

\section{Analysis of IC-OBR Type-I\label{Sec:ICOBR_ind_chn}}

In this section, we investigate the IC-OBR Type-I system described in Sec.
\ref{Sec:System Models}. We consider outer bounds and inner bounds to the
achievable rate regions for both fixed and variable OBRC bandwidth allocation.
It is noted that the results for fixed OBRC were partly presented in
\cite{onur_ISIT09}.

\subsection{Outer Bound}

In this section, we first present a general outer bound to the capacity region
of an IC-OBR in terms of multi-letter mutual informations (Proposition 1).
This bound is then specialized to a number of special cases of interest,
allowing the identification of the capacity region of IC-OBR for various scenarios.

\textbf{Proposition 1 (Outer bound for IC-OBR\ Type-I):} For fixed OBRC
bandwidth allocation, the capacity region of the IC-OBR\ Type-I is contained
within the set of rates $(R_{1},R_{2})$ satisfying
\begin{align}
&  \lim_{n\rightarrow\infty}\mbox{closure}\bigg(\bigcup_{p(x_{1}^{n},x_{2}%
^{n})=p(x_{1}^{n})p(x_{2}^{n})}\big\{(R_{1},R_{2})\text{:}\nonumber\\
&  R_{1}\leq\frac{1}{n}I(X_{1}^{n};Y_{1}^{n})+\min\left\{  \eta_{1R}%
\mathcal{C}(b_{1}^{2}P_{1R})+\eta_{2R}\mathcal{C}(b_{2}^{2}P_{2R}),\eta
_{R1}\mathcal{C}(c_{1}^{2}P_{R1})\right\} \nonumber\\
&  R_{1}\leq\frac{1}{n}I(X_{1}^{n};Y_{1}^{n}|X_{2}^{n})+\min\left\{  \eta
_{1R}\mathcal{C}(b_{1}^{2}P_{1R}),\eta_{R1}\mathcal{C}(c_{1}^{2}%
P_{R1})\right\} \nonumber\\
&  R_{2}\leq\frac{1}{n}I(X_{2}^{n};Y_{2}^{n})+\min\left\{  \eta_{1R}%
\mathcal{C}(b_{1}^{2}P_{1R})+\eta_{2R}\mathcal{C}(b_{2}^{2}P_{2R}),\eta
_{R2}\mathcal{C}(c_{2}^{2}P_{R2})\right\} \nonumber\\
&  R_{2}\leq\frac{1}{n}I(X_{2}^{n};Y_{2}^{n}|X_{1}^{n})+\min\left\{  \eta
_{2R}\mathcal{C}(b_{2}^{2}P_{2R}),\eta_{R2}\mathcal{C}(c_{2}^{2}%
P_{R2})\right\}  \big\}\bigg)\label{thm2new}%
\end{align}
where the union is taken with respect to all multi-letter input distributions
$p(x_{1}^{n})p(x_{2}^{n})$ that satisfy the power constraints $1/n\sum
_{t=1}^{n}\mbox{E}[X_{i,t}^{2}]\leq P_{i}$, $i=1,2$. With variable
OBRC\ bandwidth allocation, an outer bound is given as above but with the
union in (\ref{thm2new}) taken also with respect to all parameters $\eta
_{iR},\eta_{Ri}$, $i=1,2,$ such that $\sum_{i=1}^{2}(\eta_{iR}+\eta_{Ri}%
)=\eta$.

\emph{Proof:} Appendix A.

\subsection{Achievable Rate Region}

\label{Sec: Ach. Region}

In this section, we derive an achievable rate region for the IC-OBR Type-I. We
propose to use a rate splitting scheme similar to the standard approach for
ICs \cite{carleial} \cite{Han}. Specifically, we split the message of each
user into private and common messages, where the private message of each
source is to be decoded only by the intended destination and the common is to
be decoded at both intended and interfered destinations. However, private and
common parts are further split into two (independent) messages as follows. One
of the private message splits is sent over the IC and the other one over the
OBRC. As for the common message, both parts are sent over the IC, but one of
the two is also sent over the OBRC to the interfered destination for
interference cancellation. More specifically, we have the following four-way
split of each message $W_{i},$ $W_{i}=(W_{iR},W_{ip},W_{ic^{\prime}%
},W_{ic^{\prime\prime}})$, $i=1,2$, where: (\textit{i}) $W_{iR}\in
\lbrack1,...2^{nR_{iR}}]$ is a private message that is transmitted via the
OBRC only, directly to $D_{i}$. Notice that since the OBR has orthogonal
channels to the IC, this message is conveyed interference-free to $D_{i}$;
(\textit{ii}) $W_{ip}\in\lbrack1,...2^{nR_{ip}}]$ is a private message that is
transmitted over the IC, decoded at $D_{i}$ and treated as noise at $D_{j},$
$j\neq i$; (\textit{iii}) $W_{ic^{\prime}}\in\lbrack1,2^{nR_{ic^{\prime}}}]$
is a common message that is transmitted over the IC \textit{and} OBRC.
Specifically, the relay conveys $W_{ic^{\prime}}$ to $D_{j}$ only, $j\neq i,$
to enable interference cancellation; (\textit{iv}) $W_{ic^{\prime\prime}}%
\in\lbrack1,...2^{nR_{ic^{\prime\prime}}}]$ is a common message that is
transmitted over IC only and decoded at both destinations.

\textbf{Remark 1 (Separability and Private vs. Common Messages):} Recalling
the discussion in Sec. \ref{sec_introduction}, it is noted that the considered
transmission scheme is in general not separable, in the sense that correlated
messages are sent over the IC and OBRC. Specifically, while the private
message splits ($W_{iR},W_{ip}$) are sent separately over the two parallel
channels IC and OBRC, part of the common messages, $W_{ic^{\prime}},$ is sent
over \textit{both} IC and OBRC in order to allow interference mitigation. This
is apparently a reasonable choice for a IC-OBR Type-I, since the private
information sent on the OBRC is conveyed without interference to the intended
destination, and thus there is no need for transmission also over the IC.
Notice that transmission of the private messages over the OBRC amounts to
\textit{signal relaying,} while transmission of the common parts can be seen
as \textit{interference forwarding.}

\textbf{Proposition 2} \textbf{(Achievable Rate Region for IC-OBR\ Type-I):}
For fixed OBRC bandwidth allocation, the convex hull of the union of all rates
$(R_{1},R_{2})$ with $R_{i}=R_{ic^{\prime}}+R_{ic^{\prime\prime}}%
+R_{ip}+R_{iR}$, $i=1,2$, that satisfy the inequalities
\begin{subequations}
\begin{align}
\sum_{j\in\mathcal{S}_{1}}R_{j}  &  \leq\mathcal{C}\left(  \frac{\sum
_{j\in\mathcal{S}_{1}}k_{j1}^{2}P_{j}}{N_{1}}\right)  \text{ }\label{CS1}\\
\sum_{j\in\mathcal{S}_{2}}R_{j}  &  \leq\mathcal{C}\left(  \frac{\sum
_{j\in\mathcal{S}_{2}}k_{j2}^{2}P_{j}}{N_{2}}\right)  \text{ }\label{CS2}\\
R_{1c^{\prime}}+R_{1R}  &  \leq\eta_{1R}\mathcal{C}(b_{1}^{2}P_{1R}%
)\label{C1Rach}\\
R_{2c^{\prime}}+R_{2R}  &  \leq\eta_{2R}\mathcal{C}(b_{2}^{2}P_{2R}%
)\label{C2Rach}\\
R_{2c^{\prime}}+R_{1R}  &  \leq\eta_{R1}\mathcal{C}(c_{1}^{2}P_{R1}%
)\label{CR1ach}\\
R_{1c^{\prime}}+R_{2R}  &  \leq\eta_{R2}\mathcal{C}(c_{2}^{2}P_{R2}%
),\label{ach_reg_end}%
\end{align}
provides an achievable rate region for the Gaussian IC-OBR Type-I, where
conditions (\ref{CS1})-(\ref{CS2}) must hold for all subsets $\mathcal{S}%
_{1}\subseteq\mathcal{T}_{1}=\{1c,1p,2c^{\prime\prime}\}$ and $\mathcal{S}%
_{2}\subseteq\mathcal{T}_{2}=\{2c,2p,1c^{\prime\prime}\}$, \textit{except
}$\mathcal{S}_{1}=\{2c^{\prime\prime}\}$ and $\mathcal{S}_{2}=\{1c^{\prime
\prime}\};$ and we define $R_{ic}=R_{ic^{\prime}}+R_{ic^{\prime\prime}},$
$N_{1}=a_{21}^{2}P_{2p}+1$, $N_{2}=a_{12}^{2}P_{1p}+1$, and the parameters
$k_{j1}=1$, $k_{j2}=a_{12}$ if $j\in\{1c,1p\}$, and $k_{j1}=a_{21}$, $k_{j2}=1
$ if $j\in\{2c,2p\}$. Moreover, we use the convention $P_{jc^{\prime\prime}%
}=P_{jc}$, $j=1,2$, and the power allocations must satisfy the power
constraints $P_{1c}+P_{1p}\leq P_{1}$ and $P_{2c}+P_{2p}\leq P_{1}$. With
variable OBRC\ bandwidth allocation, rates (\ref{CS1})-(\ref{ach_reg_end}) can
be evaluated for all bandwidth allocations satisfying $\sum_{i=1}^{2}%
(\eta_{iR}+\eta_{Ri})=\eta$.

\emph{Proof:} Appendix B.

\subsection{Review of Capacity Results}

Here we review the main capacity results for IC-OBR Type-I with fixed and
variable bandwidth allocation that will be detailed in the following. We are
interested in assessing under which conditions signal relaying or a
combination of signal relaying and interference forwarding attain optimal
peformance. According to the discussion above, optimality of signal relaying
alone implies that separable operation on the IC-OBR is optimal, whereas, if
interference forwarding is needed, a separable scheme (at least in the class
of strategies we considered) is not sufficient to attain optimal performance.
The derived optimality conditions set constraints on both the IC and
OBRC\ channel gains. All results are obtained from the outer bound of
Proposition 1 and the achievable rate region of Proposition 2.

Starting with \textit{fixed} bandwidth allocation, we first classify IC-OBR
systems\ on the basis of the channel conditions on the OBRC. Specifically, we
distinguish two regimes:
\end{subequations}
\begin{itemize}
\item \textit{Relay-to-destinations (R-to-D) bottleneck regime}: In this
regime, along the signal paths $S_{i}-R-D_{i}$ on the OBRC,\ for $i=1,2,$ the
relay to destinations links form the performance bottleneck. In other words,
the channel from source $S_{i}$ to relay $R$ is better than the one from relay
$R$ to destination $D_{i}$ for $i=1,2.$ Specifically, we have the conditions
\begin{equation}
\eta_{R1}\mathcal{C}(c_{1}^{2}P_{R1})\leq\eta_{1R}\mathcal{C}(b_{1}^{2}%
P_{1R})\text{ and }\eta_{R2}\mathcal{C}(c_{2}^{2}P_{R2})\leq\eta
_{2R}\mathcal{C}(b_{2}^{2}P_{2R}).\label{bottleneck_conditions}%
\end{equation}

\item \textit{Excess rate regime}: In this regime, the bottleneck condition is
not satisfied and we assume that
\begin{equation}
\eta_{R1}\mathcal{C}(c_{1}^{2}P_{R1})\geq\eta_{1R}\mathcal{C}(b_{1}^{2}%
P_{1R})\text{ and }\eta_{R2}\mathcal{C}(c_{2}^{2}P_{R2})\leq\eta
_{2R}\mathcal{C}(b_{2}^{2}P_{2R}).\label{ex_rate_conditions}%
\end{equation}
A symmetric regime can also be equivalently considered where the second
inequality in (\ref{bottleneck_conditions}) is violated rather than the first
as in (\ref{ex_rate_conditions}). For reasons that will be made clear below,
under condition (\ref{ex_rate_conditions}), we define the \textit{excess rate}
from $S_{2}$ to $D_{1}$ on the OBRC as
\begin{equation}
R_{ex_{21}}=\min\left\{  \eta_{R1}\mathcal{C}(c_{1}^{2}P_{R1})-\eta
_{1R}\mathcal{C}(b_{1}^{2}P_{1R}),\text{ }\eta_{2R}\mathcal{C}(b_{2}^{2}%
P_{2R})-\eta_{R2}\mathcal{C}(c_{2}^{2}P_{R2})\right\}  .\label{ex_rate}%
\end{equation}

\end{itemize}

We first show that in the R-to-D bottleneck regime, \textit{irrespective of
the channel gains over the IC}, signal relaying alone, and thus
\textit{separable} operation, is optimal (Proposition 3). In particular, in
this regime, it is optimal to transmit only \textit{private} information over
the OBRC at the maximum rate $R_{iR}$ on each link $S_{i}-R-D_{i}.$ Since this
rate is limited by (\ref{bottleneck_conditions}), we have $R_{iR}=\min
\{\eta_{Ri}\mathcal{C}(c_{i}^{2}P_{Ri}),\eta_{iR}\mathcal{C}(b_{i}^{2}%
P_{iR})\}=\eta_{iR}\mathcal{C}(b_{i}^{2}P_{iR})$ for $i=1,2.$ Notice every bit
of private rate $R_{iR}$ directly adds one bit to the overall achievable rate
for the pair $S_{i}-D_{i},$ since it consists of independent information sent
directly to the destination $D_{i}$ (see also \cite{wei yu}).

We then turn to the excess rate regime (\ref{ex_rate_conditions}) (symmetric
results clearly holds by swapping indices 1 and 2). Assume that both sources
transmit private bits over the OBRC at their maximum rate $R_{iR}.$ Given
condition (\ref{ex_rate_conditions}), this leads to $R_{1R}=\min\{\eta
_{R1}\mathcal{C}(c_{1}^{2}P_{R1}),\eta_{1R}\mathcal{C}(b_{1}^{2}P_{1R}%
)\}=\eta_{1R}\mathcal{C}(b_{1}^{2}P_{1R})$ and $R_{2R}=\min\{\eta
_{R2}\mathcal{C}(c_{2}^{2}P_{R2}),\eta_{2R}\mathcal{C}(b_{2}^{2}P_{2R}%
)\}=\eta_{R2}\mathcal{C}(c_{2}^{2}P_{R2}).$ As said, these bits directly
contribute to the overall achievable rates for the two source-destination
pairs. Having allocated such rates for signal relaying, one is left with the
\textit{excess rate} (\ref{ex_rate}) on the OBRC\ between source $S_{2}$ and
destination $D_{1}.$ This excess rate can be potentially used for interference
relaying. We show that the discussed rate allocation over the OBRC with
interference forwarding at the excess rate is optimal in two specific regimes
of channel gains over the IC.

More precisely, we prove that signal relaying and interference forwarding (via
the excess rate) is optimal in case destination $D_{2}$ is in either
\textit{strong} interference conditions on the IC (i.e., $a_{12}\geq1)$ in
Proposition 4 or \textit{weak} interference conditions (i.e., $a_{12}<1)$ in
Proposition 5. In both cases, the excess rate (\ref{ex_rate}) on the
OBRC\ link $S_{2}-R-D_{1},$ thanks to \textit{interference forwarding}, has
the effect of pushing destination $D_{1}$ in a "strong" interference regime
over the IC-OBR, so that $D_{1}$ can decode source $S_{2}$'s messages without
loss of optimality. Proposition 4 is thus akin to the standard
strong-interference capacity region result for the Gaussian IC \cite{sato},
while Proposition 5 is akin to the mixed-interference sum-capacity result for
the Gaussian IC \cite{khandani}\cite{Tuninetti}. It is emphasized that here
strong and weak interference conditions are attained on the IC-OBR and not on
the IC alone (destination $D_{1}$ is generally not in the strong interference
regime on the IC alone). We also show that, if destination $D_{1}$ is already
in sufficiently strong interference conditions on the IC alone, interference
forwarding over the OBRC\ becomes unnecessary (see Remark 3 and 4).

We then consider variable bandwith allocation over the OBRC and show that
under appropriate conditions, allocating all the bandwidth to the better path
for signal relaying is optimal, while this is not the case under more general
assumptions and interference forwarding may be useful.

\subsection{Capacity Results for Fixed OBRC Bandwidth Allocation}

In this section, we detail the capacity results reviewed above for fixed OBRC
bandwidth allocation.

\subsubsection{R-to-D Bottleneck Regime}

We first focus on the R-to-D bottleneck regime (\ref{bottleneck_conditions}).
The proposition is expressed in terms of the capacity region $\mathcal{C}%
_{IC}$ of a standard IC, which is generally unknown in single-letter
formulation apart from special cases.

\textbf{Proposition 3 (R-to-D bottleneck regime}$)$\textbf{:} For a
IC-OBR\ Type-I with fixed OBRC\ bandwidth allocation, under condition
(\ref{bottleneck_conditions}), the capacity region $\mathcal{C}$ is given by
the capacity region $\mathcal{C}_{IC}$ of the IC, enhanced by ($\eta
_{R1}\mathcal{C}(c_{1}^{2}P_{R1}),\eta_{R2}\mathcal{C}(c_{2}^{2}P_{R2}))$
along the individual rates as
\begin{equation}
\mathcal{C}=\{(R_{1},R_{2})\text{: }(R_{1}-R_{1}^{\prime},R_{2}-R_{2}^{\prime
})\in\mathcal{C}_{IC}\},\nonumber
\end{equation}
for $R_{1}^{\prime}\leq\eta_{R1}\mathcal{C}(c_{1}^{2}P_{R1})$, $R_{2}^{\prime
}\leq\eta_{R2}\mathcal{C}(c_{2}^{2}P_{R2})$. Equivalently, the capacity region
$\mathcal{C}$ is given by
\begin{align}
\mathcal{C} &  =\lim_{n\rightarrow\infty}\mbox{closure}\bigg(\bigcup
_{p(x_{1}^{n},x_{2}^{n})=p(x_{1}^{n})p(x_{2}^{n})}\big\{(R_{1},R_{2}%
)\text{:}\nonumber\\
&  R_{1}\leq\frac{1}{n}I(X_{1}^{n};Y_{1}^{n})+\eta_{R1}\mathcal{C}(c_{1}%
^{2}P_{R1})\nonumber\\
&  R_{2}\leq\frac{1}{n}I(X_{2}^{n};Y_{2}^{n})+\eta_{R2}\mathcal{C}(c_{2}%
^{2}P_{R2})\big\}\bigg),
\end{align}
where the union is taken with respect to the input distribution $p(x_{1}%
^{n})p(x_{2}^{n})$ that satisfies the power constraints $1/n\sum_{t=1}%
^{n}E[X_{i,t}^{2}]\leq P_{i},$ $i=1,2.$ The capacity region is achieved by
signal relaying only.

\emph{Proof:} The converse follows immediately from Proposition 1 with the
conditions $\eta_{Ri}\mathcal{C}(c_{i}^{2}P_{Ri})\leq\eta_{iR}\mathcal{C}%
(b_{i}^{2}P_{iR})$, $i=1,2$ and by Ahlswede's multi-letter characterization of
the interference channel capacity region \cite{ahlswede}. Achievability
follows by sending independent messages $(W_{1R},W_{2R}) $ in the notation of
Proposition 2, of rates $\eta_{R1} \mathcal{C}(c_{1}^{2}P_{R1})$ and
$\eta_{R2} \mathcal{C}(c_{2}^{2}P_{R2})$ from sources $S_{1}$ and $S_{2}$,
respectively, on the OBRC, and then using the Gaussian IC as a regular IC,
stripped of the OBRC. $\Box$

\textbf{Remark 2: }Proposition 3 does not impose any conditions on the IC.
Therefore, in any scenario where a single-letter capacity region is known for
the IC, the single-letter capacity region immediately carries over to the
IC-OBR Type-I with (\ref{bottleneck_conditions})$.$ For instance, we can
obtain a single-letter capacity region expression for an IC-OBR Type-I in the
strong interference regime ($a_{21}\geq1$ and $a_{12}\geq1)$ \cite{Han}
\cite{sato} or the sum-capacity in the noisy or mixed interference regime
\cite{Kramer_int_cap}\cite{Veeravalli_int_cap}\cite{khandani}\cite{Tuninetti},
as long as (\ref{bottleneck_conditions}) holds. Moreover, both Proposition 2
and 3 apply also to a general discrete memoryless IC-OBR Type-I (with the
caveat of eliminating the power constraint).

\subsubsection{Excess Rate Regime}

\label{Sec:Type_I_Int_Forw} We next investigate the capacity region for the
excess rate regime (\ref{ex_rate_conditions}). Proposition 4 is formulated
under conditions akin to strong interference conditions for standard ICs
\cite{sato}, while Proposition 5 holds for assumptions akin to the mixed
interference regime \cite{khandani}\cite{Tuninetti}.

\textbf{Proposition 4 (Excess rate regime, Strong interference):} For an
IC-OBR Type-I with fixed OBRC\ bandwidth allocation, under conditions
(\ref{ex_rate_conditions}), if $a_{12}\geq1,$ we have that : (\textit{i}) If
$R_{ex_{21}}\geq\max\left\{  0,\mathcal{C}(a_{12}^{2}P_{1}+P_{2}%
)-\mathcal{C}(P_{1}+a_{21}^{2}P_{2})\right\}  $, the following inequalities
characterize the capacity region,
\begin{subequations}
\begin{align}
R_{1} &  \leq\mathcal{C}\left(  P_{1}\right)  +\eta_{1R}\mathcal{C}(b_{1}%
^{2}P_{1R})\label{R1th4}\\
R_{2} &  \leq\mathcal{C}\left(  P_{2}\right)  +\eta_{R2}\mathcal{C}(c_{2}%
^{2}P_{R2})\label{R2th4}\\
R_{1}+R_{2} &  \leq\mathcal{C}\left(  a_{12}^{2}P_{1}+P_{2}\right)  +\eta
_{1R}\mathcal{C}(b_{1}^{2}P_{1R})+\eta_{R2}\mathcal{C}(c_{2}^{2}%
P_{R2});\label{Rtotth4}%
\end{align}
(\textit{ii}) If $0\leq R_{ex_{21}}=\eta_{R1}\mathcal{C}(c_{1}^{2}P_{R1}%
)-\eta_{1R}\mathcal{C}(b_{1}^{2}P_{1R})\leq\mathcal{C}(a_{12}^{2}P_{1}%
+P_{2})-\mathcal{C}(P_{1}+a_{21}^{2}P_{2})$, the following conditions
characterize the capacity region
\end{subequations}
%\begin{subequations}
\label{Rtotth5}%
\begin{align}
R_{1} &  \leq\mathcal{C}\left(  P_{1}\right)  +\eta_{1R}\mathcal{C}(b_{1}%
^{2}P_{1R})\label{R1th5}\\
R_{2} &  \leq\mathcal{C}\left(  P_{2}\right)  +\eta_{R2}\mathcal{C}(c_{2}%
^{2}P_{R2})\label{R2th5}\\
R_{1}+R_{2} &  \leq\mathcal{C}\left(  P_{1}+a_{21}^{2}P_{2}\right)  +\eta
_{R1}\mathcal{C}(c_{1}^{2}P_{R1})+\eta_{R2}\mathcal{C}(c_{2}^{2}P_{R2}).
\end{align}
In both cases (\textit{i}) and (\textit{ii}) the capacity region is achieved
by joint signal relaying and interference forwarding.

\textbf{Proof:} Appendix C.

\textbf{Remark 3: }Under the conditions of Proposition 4, $D_{2}$ is in the
strong interference regime ($a_{12}\geq1$) on the IC and, as a consequence, it
can be proved that transmission of only common information by $S_{1}$ over the
IC\ is optimal (separable operation). Moreover, the excess rate guarantees
that also $D_{1}$ is essentially driven in the strong interference regime in
the IC-OBR\footnote{Notice that we do not necessarily have strong interference
($a_{21}\geq1)$ on the IC in case (\textit{i}) but we do in case
(\textit{ii})$.$ }$.$ As a result, $S_{2}$ can also transmit only common
information with the caveat that part of it will be sent also over the OBRC
(with rate $R_{2c^{\prime}},$ non-separable operation). To be more specific,
the assumptions in Proposition 4 encompass two different situations. In case
(\textit{i}), the sum-rate bound (\ref{Rtotth4}) to receiver $D_{2}$ sets the
sum-rate performance bottleneck, and interference forwarding is performed from
$S_{2}$ to $D_{1}$ with rate equal to $R_{2c^{\prime}}=(\mathcal{C}(a_{12}%
^{2}P_{1}+P_{2})-\mathcal{C}(P_{1}+a_{21}^{2}P_{2}))^{+}$. In particular, if
$(a_{12}^{2}-1)P_{1}+(1-a_{21}^{2})P_{2}\leq0$, which implies that $a_{21}$ is
large enough, we can set $R_{2c^{\prime}}=0$ and there is no need for
interference forwarding$.$ On the other hand, for $(a_{12}^{2}-1)P_{1}+(1-a_{21}^{2})P_{2}>0$, which can hold even when $a_{21}<1$,
the optimal scheme necessitates interference forwarding, i.e $R_{2c^{\prime}}>0$. In case (\textit{ii}) the sum-rate bound
(\ref{thm4_2}) for $D_{1}$ is always more restrictive than (\ref{Rtotth4}) for
$D_{2}$ (see Appendix C) in terms of the sum-rate and interference forwarding
is performed with $R_{2c^{\prime}}=R_{ex_{21}}=\eta_{R1}\mathcal{C}(c_{1}%
^{2}P_{R1})-\eta_{1R}\mathcal{C}(b_{1}^{2}P_{1R}).$

\begin{figure}[t]
%\end{subequations}
\begin{center}
\includegraphics[width=10 cm]{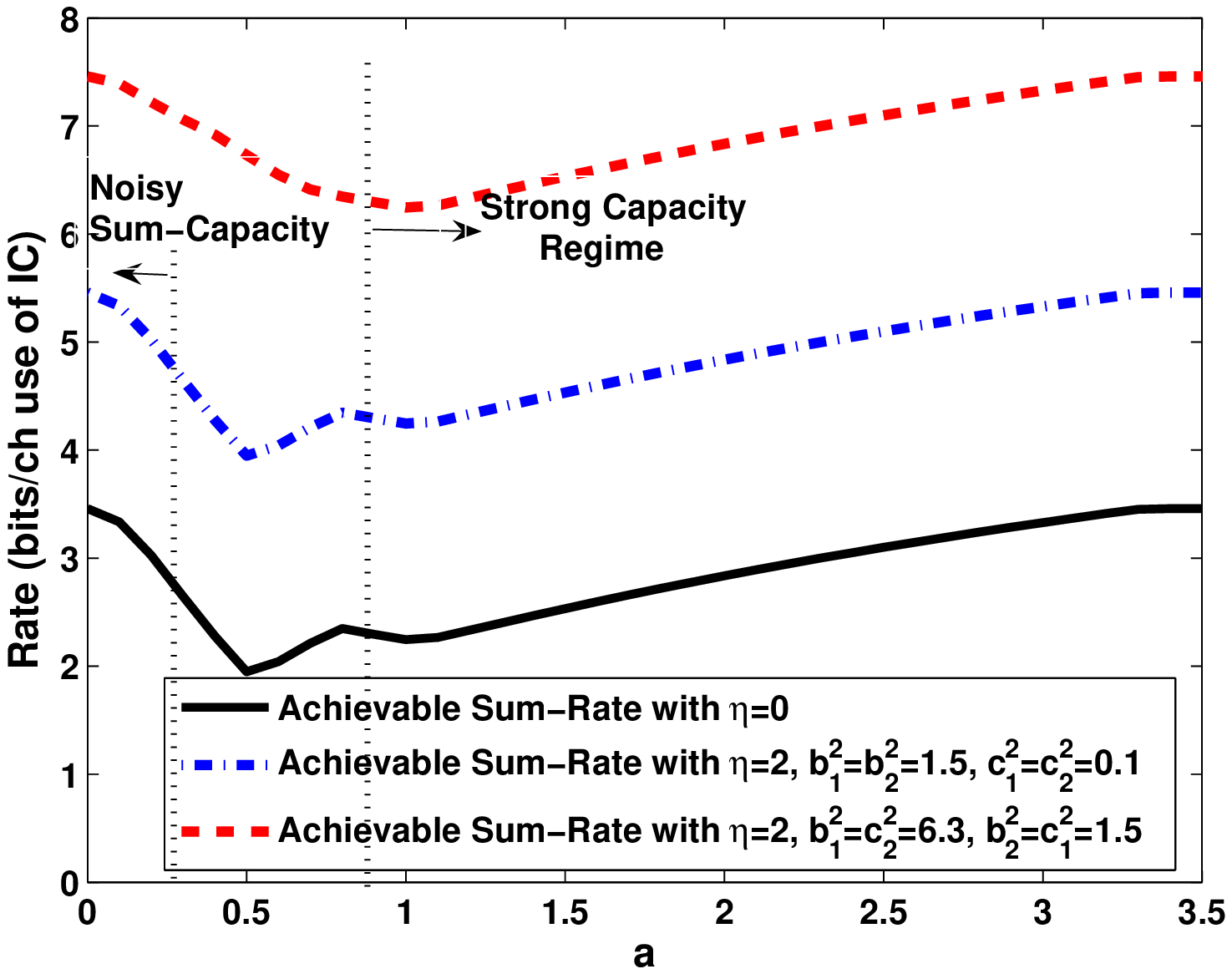}
\end{center}
\caption{Achievable sum-rate from Proposition 2 for IC-OBR\ Type-I with
symmetric IC, for fixed OBRC\ bandwidth allocation versus $a$ for various OBRC
link capacities, namely \emph{i)} $\eta=0$ (no relay), \emph{ii)} $\eta
_{1R}=\eta_{2R}=\eta_{R1}=\eta_{R2}=1$, $b_{1}^{2}=b_{2}^{2}=1.5$, $c_{1}%
^{2}=c_{2}^{2}=0.3$, (which satisfies the conditions in Proposition 3), and
\emph{iii)} $\eta_{1R}=\eta_{2R}=\eta_{R1}=\eta_{R2}=1$, $b_{2}^{2}=c_{1}%
^{2}=6.3$, $b_{1}^{2}=c_{2}^{2}=1.5$ (which satisfies the conditions in
Proposition 4). All powers are set to $10$ dB.}
\label{Fig:ach_1}
\end{figure}

In Fig. \ref{Fig:ach_1}, we show the maximum achievable sum-rate of
Proposition 2 for different configurations of the OBRC channel gains and
bandwidths. We have a symmetric IC with $P=P_{1}=P_{2}=10$dB and
$a_{21}=a_{12}=a$. For comparison, we show the no relay case $\eta_{1R}%
=\eta_{2R}=\eta_{R1}=\eta_{R2}=0$. Moreover, we first consider a scenario
where the relay-to-destination links form bottleneck with respect to
source-to-relay links, i.e. $\eta_{1R}\mathcal{C}(b_{1}^{2}P_{1R})=\eta
_{2R}\mathcal{C}(b_{2}^{2}P_{2R})=2,\eta_{R1}\mathcal{C}(c_{1}^{2}P_{R1}%
)=\eta_{R2}\mathcal{C}(c_{2}^{2}P_{R2})=1,$ thus falling within the
assumptions of Proposition 3. It can be seen that the sum-rate increases by
$2$ bits/ch use of IC for all values of $a$. Moreover, from\ Proposition 3, it
is known that in the \emph{noisy }\cite{Veeravalli_int_cap} $\left(
a(1+10a^{2})\leq0.5, \mbox{ i.e, }a\leq0.28\right) $ and \emph{strong }%
($a\geq1$) \cite{sato} interference regimes, the sum-rate shown in the figure
is the sum-capacity. Finally, we consider a situation with $\eta
_{2R}\mathcal{C}(b_{2}^{2}P_{2R})=\eta_{R1}\mathcal{C}(c_{1}^{2}P_{R1}%
)=3,\eta_{1R}\mathcal{C}(b_{1}^{2}P_{1R})=\eta_{R2}\mathcal{C}(c_{2}^{2}%
P_{R2})=2,$ which falls under the conditions of Proposition 4 for $a\geq1$. As
stated in the Proposition, for $a\geq1$, the sum-rate shown in the
sum-capacity is $4$ bits/channel use of IC larger than the reference case of
zero OBRC capacities.

\textbf{Proposition 5 (Excess rate regime, Mixed interference)}: For an IC-OBR
Type-I with fixed OBRC\ bandwidth allocation and the condition $a_{12}<1$, we
have that if $R_{ex_{21}}\geq\max\left\{  0,\mathcal{C}(P_{1})+\mathcal{C}%
\left(  \frac{P_{2}}{1+a_{12}^{2}P_{1}}\right)  -\mathcal{C}\left(
P_{1}+a_{21}^{2}P_{2}\right)  \right\}  $, the following condition
characterizes the sum capacity
\[
R_{1}+R_{2}\leq\mathcal{C}\left(  P_{1}\right)  +\mathcal{C}\left(
\frac{P_{2}}{1+a_{12}^{2}P_{1}}\right)  +\eta_{1R}\mathcal{C}(b_{1}^{2}%
P_{1R})+\eta_{R2}\mathcal{C}(c_{2}^{2}P_{R2}),
\]
which is achieved by joint signal relaying and interference forwarding.

\emph{Proof: }Appendix D.

\textbf{Remark 4:} Since $D_{2}$ observes weak interference, the optimal
coding strategy for $S_{1}$ is to transmit private messages only, separably
over both IC and OBRC. As for Proposition 4, interference forwarding
essentially drives $D_{2}$ in a strong interference regime (even though we do
not have $a_{21}\geq1$ in general). Therefore, $S_{2}$ transmits only common
messages in a non-separable way over both IC\ and OBRC via interference
forwarding. It is also noted, similar to Remark 3, that if the interference at
$D_{2}$ is strong enough on the IC, and in particular, we have%
\begin{equation}
a_{21}\geq\sqrt{\frac{1+P_{1}}{1+a_{12}^{2}P_{1}}},\label{strong_IC}%
\end{equation}
then it can be shown that interference forwarding is not necessary and one can
set $R_{2c^{\prime}}=0$.

\subsection{Capacity Results for OBRC Variable Bandwidth Allocation}

\label{Sec:TypeI_var_bw}

In this part, we discuss optimality of the considered strategies under
variable OBRC bandwidth allocation. The next proposition shows that, thanks to
the ability to allocate bandwidth among the source-to-relay and
relay-to-destination channels, under certain conditions bandwidth can be
allocated only to the \emph{best} channels and in a way that only signal
relaying is used.

\textbf{Proposition 6: }Consider an IC-OBR Type-I channel with variable
OBRC\ bandwidth allocation and $a_{12}=a_{21}\geq1$, $P_{1}=P_{2}$,
$\mathcal{C}(b_{1}^{2}P_{1R})=\mathcal{C}(c_{1}^{2}P_{R1})$, $\mathcal{C}%
(b_{1}^{2}P_{1R})\geq\mathcal{C}(b_{2}^{2}P_{2R})$ and $\mathcal{C}(c_{1}%
^{2}P_{R1})\geq2\mathcal{C}(c_{2}^{2}P_{R2})$. The sum capacity is given by
\[
R_{1}+R_{2}\leq\mathcal{C}\left(  a_{12}^{2}P_{1}+P_{2}\right)  +\frac{\eta
}{2}\mathcal{C}(c_{1}^{2}P_{R1})
\]
and is achieved by signal relaying and transmitting only common information on
the IC.

\emph{Proof:} Appendix E.

\textbf{Remark 5:} Proposition 6 states that for a symmetric and strong IC,
when the path $S_{1}-R-D_{1}$ is better than $S_{2}-R-D_{2}$ (see conditions
on $b_{i}$ and $c_{i}),$ under appropriate conditions, it is optimal to
allocate all the bandwidth $\eta$ to the better path to perform signal
relaying. Signal relaying increases the sum-rate by the number of bits carried
on the OBRC links, namely $\frac{\eta}{2}\mathcal{C}(c_{1}^{2}P_{R1}).$ Notice
that it is possible to generalize Proposition 6 to arbitrary $b_{1}$ and
$c_{1}$, i.e. not necessarily restricting the channels to satisfy the
condition $\mathcal{C}(b_{1}^{2}P_{1R})=\mathcal{C}(c_{1}^{2}P_{R1})$.
However, here we have focused on this simple case, to illuminate more clearly
the gist of the main result.

\begin{figure}[t]
\begin{center}
\includegraphics[width=10 cm]{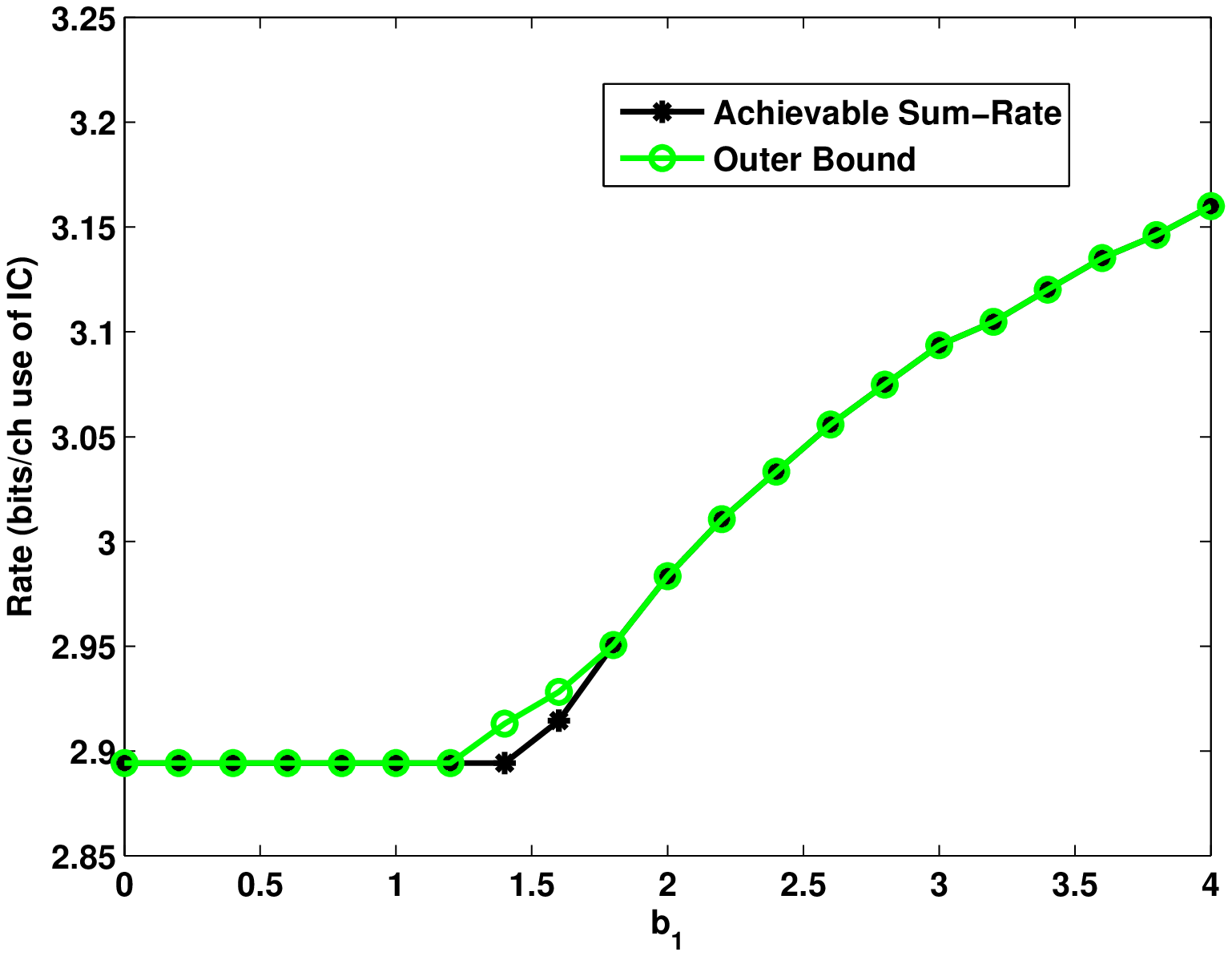}
\end{center}
\caption{Achievable sum-rate (from Proposition 2, (\ref{ach_rate_sum_var})
with signal relaying (only common information transmission over the IC), and
outer bound (from Proposition 1, (\ref{TypeI_OB})) versus $b_{1}$ for
IC-OBR\ Type-I with variable OBRC\ bandwidth allocation ($\eta=1$,
$a_{21}=a_{12}=2$, all powers equal to $10$ dB, $c_{1}=3,c_{2}=b_{2}=2)$.}%
\label{Fig:Ach_vs_out_b}%
\end{figure}

To obtain further insight into the result of Proposition 6, in Fig.
\ref{Fig:Ach_vs_out_b} we compare the sum-rate achievable by transmitting only
common information on the IC and using signal relaying (with optimized
bandwidth allocation), obtained from Proposition 2 (see
(\ref{ach_rate_sum_var}) in Appendix E), and the upper bound obtained from
Proposition 1 (see (\ref{TypeI_OB}) in Appendix E). Parameters are fixed as
$\eta=1$, $a_{21}=a_{12}=2$, $c_{1}=3,c_{2}=b_{2}=2,$ all powers equal to $10$
dB and $b_{1}$ is varied. Numerical results show for this example that the
achievable rate matches the upper bound for most of the channel gains (except
$1.2\leq b_{1}\leq1.8$).

\begin{figure}[t]
\begin{center}
\includegraphics[width=10 cm]{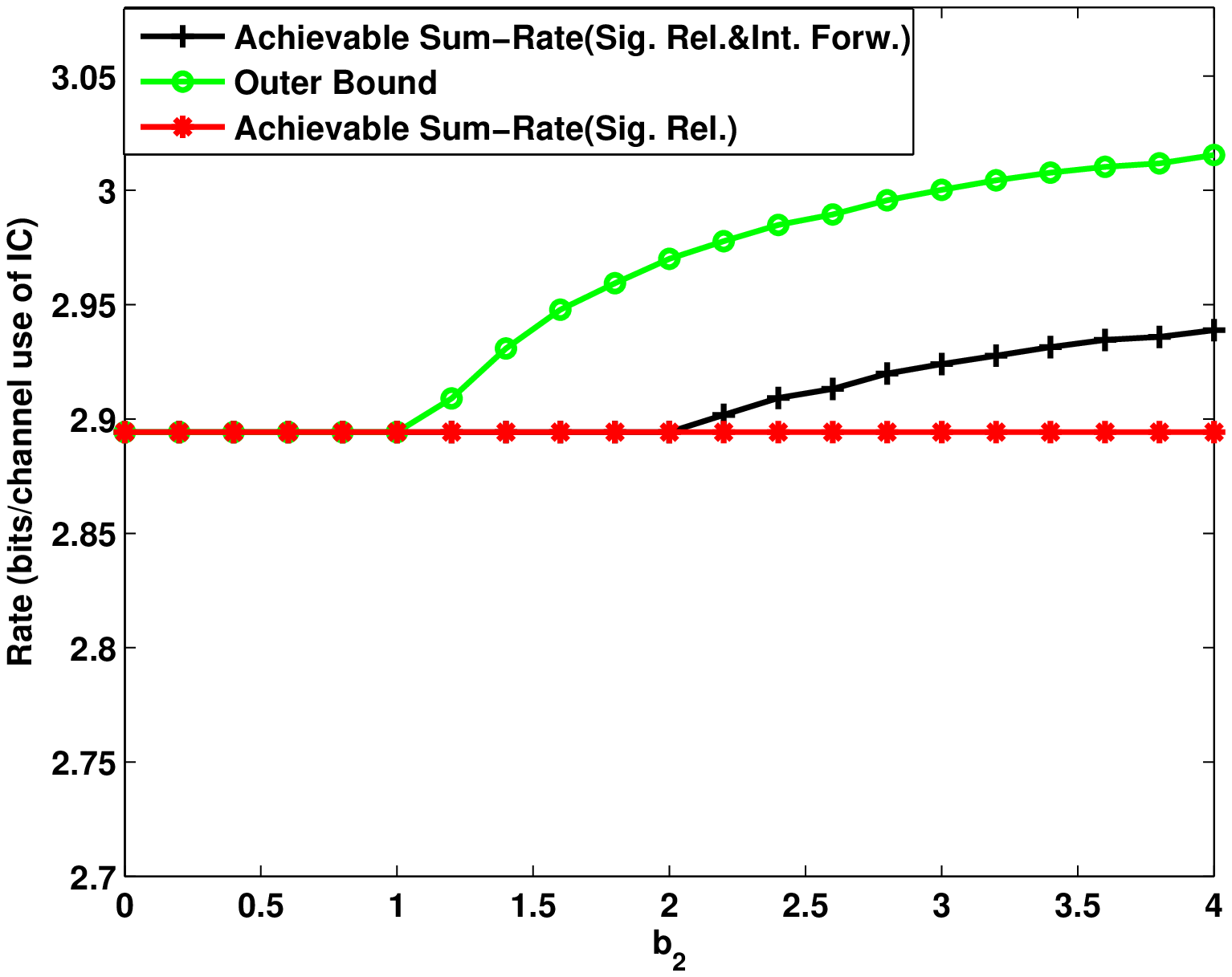}
\end{center}
\caption{Achievable sum-rate (from Proposition 4, (\ref{ach_reg_thm4}%
)-(\ref{thm4_2})) with signal relaying and interference forwarding (only
common information transmission over the IC) and outer bound (from Proposition
1, (\ref{TypeI_OB})) versus $b_{2}$ for IC-OBR\ Type-I with variable
OBRC\ bandwidth allocation $\sum_{i=1}^{2}\eta_{iR}+\eta_{Ri}=\eta,$ $\eta=1$,
$a_{12}=3$, $a_{21}=2$, all powers equal to $10$ dB, $b_{1}=1.5$, $c_{1}=3$,
and $c_{2}=1$.}%
\label{Fig:TypeI_varBW}%
\end{figure}

The result in Proposition 6 begs the question as to whether interference
forwarding can be ever useful in the case of variable bandwidth allocation.
The following example answers this question in the affirmative. Consider an
IC-OBR Type-I channel with $b_{1}=1.5$, $c_{1}=3$, $c_{2}=1$, $a_{12}=3$,
$a_{21}=2,$ $\eta=1$ and all transmission powers set to $10$ dB. Here, $b_{2}$
is varied. Fig. \ref{Fig:TypeI_varBW} shows the achievable sum-rate obtained
from Proposition 4 (which was derived from Proposition 2) by assuming
transmission of common messages only over the IC\ and either signal relaying
only ($(W_{ic^{\prime\prime}},W_{iR})$, $i=1,2$) or both signal relaying and
interference forwarding ($(W_{1c^{\prime\prime}},W_{1R})$,$(W_{2c^{\prime}%
},W_{2c^{\prime\prime}},W_{2R})$) (see (\ref{ach_reg_thm4})-(\ref{thm4_2}) in
Appendix C) along with optimized bandwidth allocation ($\eta_{iR},\eta_{Ri}),$
$i=1,2$. Also shown is the outer bound from Proposition 1, given in
(\ref{TypeI_OB}) for the channel parameters at hand. For $b_{2}\geq c_{2}=1$
and for fixed and equal bandwidth allocations, $\eta_{1R}=\eta_{R1}=\eta
_{2R}=\eta_{R2}$, the excess rate $R_{ex_{21}}$ is positive and interference
forwarding is potentially useful as discussed in Propositions 4. For the
variable bandwidth case, Fig. \ref{Fig:TypeI_varBW} shows similarly that
interference-forwarding is instrumental in improving the achievable sum-rate
for $b_{2}\geq2$. However, for $b_{2}<2$, interference forwarding is not needed.

\section{Analysis of IC-OBR Type-II}

\label{Sec:IC-OBR Type II} In this section, we investigate the IC-OBR Type-II
channel described in Sec. \ref{Sec:System Models}. This channel differs from
IC-OBR\ Type-I in that on the OBRC: (\textit{i}) The signals transmitted from
$S_{1}$ and $S_{2}$ are superimposed at the relay; (\textit{ii}) Relay
broadcasts to the destinations. These aspects allow novel, and more general,
transmission strategies at sources and relay, thus making the design of
optimal schemes more complex than on the IC-OBR\ Type-I. This is reflected by
the results presented below that encompass scenarios and techniques that have
a counterpart in the IC-OBR\ Type-I analysis and others that stand out as
specific to IC-OBR\ Type-II.

To simplify the analysis, we still consider the same class of strategies
described in Sec. \ref{Sec: Ach. Region} for source and destinations. Namely,
we assume a four-way message split into two private and two common parts, such
that, as discussed in Remark 1, private signals are transmitted in a separable
way over IC and OBRC, while\ one of the common messages is sent over both IC
and OBRC (and, hence, transmitted in a non-separable way). Even within this
class of strategies, the variety of possible approaches is remarkable. For
instance, the relay may perform Decode-and-Forward (DF) or different flavors
of Compress-and-Forward (CF) \cite{Razaghi}\cite{noisy} and the sources may
encode by using random codes or structured (e.g., lattice) codes
\cite{wilson_lattice}\cite{nam_lattice}. For this reason, unlike IC-OBR
Type-I, we will not give a general achievable rate region but rather focus on
specific conditions under which optimality of certain design choices can be assessed.

We first give an outer bound on the capacity region for IC-OBR\ Type-II.
Then, we investigate conditions under which signal relaying (and thus
separable operation, as per the discussion above) or joint signal relaying/
interference forwarding (and thus non-separable strategies) are optimal.

\subsection{Outer Bound}

In the following, we give an outer bound that is the counterpart of
Proposition 1 for IC-OBR Type-II.

\textbf{Proposition 7 (Outer Bound for IC-OBR Type-II):} For an
IC-OBR\ Type-II with fixed OBRC\ bandwidth allocation, the capacity region is
included in the following region
\begin{subequations}
\label{R2outc_end}%
\begin{align}
&  \lim_{n\rightarrow\infty}\mbox{closure}\bigg(\bigcup_{p(x_{1}^{n},x_{2}%
^{n})=p(x_{1}^{n})p(x_{2}^{n})}\big\{(R_{1},R_{2})\text{:}\nonumber\\
R_{1} &  \leq\frac{1}{n}I(X_{1}^{n};Y_{1}^{n})+\eta_{MAC}\mathcal{C}\left(
b_{1}^{2}P_{1R}+b_{2}^{2}P_{2R}\right)  \label{R1out1}\\
R_{1} &  \leq\frac{1}{n}I(X_{1}^{n};Y_{1}^{n}|X_{2}^{n})+\eta_{MAC}%
\mathcal{C}\left(  b_{1}^{2}P_{1R}\right)  \label{R1out2}\\
R_{1} &  \leq\frac{1}{n}I(X_{1}^{n};Y_{1}^{n}|X_{2}^{n})+\eta_{BC}%
\mathcal{C}\left(  c_{1}^{2}P_{R}\right)  \label{R1_p8}\\
R_{2} &  \leq\frac{1}{n}I(X_{2}^{n};Y_{2}^{n})+\eta_{MAC}\mathcal{C}\left(
b_{1}^{2}P_{1R}+b_{2}^{2}P_{2R}\right)  \label{R2out1}\\
R_{2} &  \leq\frac{1}{n}I(X_{2}^{n};Y_{2}^{n}|X_{1}^{n})+\eta_{MAC}%
\mathcal{C}\left(  b_{2}^{2}P_{2R}\right)  \label{R2out2}\\
R_{2} &  \leq\frac{1}{n}I(X_{2}^{n};Y_{2}^{n}|X_{1}^{n})+\eta_{BC}%
\mathcal{C}\left(  c_{2}^{2}P_{R}\right)  \big\}\bigg)\label{R2_p8}%
\end{align}
where the union is taken with respect to multi-letter input distributions
$p(x_{1}^{n})p(x_{2}^{n})$ that satisfy the power constraints $1/n\sum
_{t=1}^{n}\mbox{E}[X_{i,t}^{2}]\leq P_{i}$, $i=1,2$. Moreover, if $a_{12}%
\leq1$ and $c_{1}\geq c_{2},$ the capacity region is included in a region
given as above but with
\end{subequations}
\begin{align}
R_{1}  & \leq\frac{1}{n}I(X_{1}^{n};Y_{1}^{n}|X_{2}^{n})+\eta_{BC}%
\mathcal{C}(c_{1}^{2}\xi P_{R})\label{R1out3}\\
R_{2}  & \leq\frac{1}{n}I(X_{2}^{n};Y_{2}^{n})+\eta_{BC}\mathcal{C}\left(
\frac{c_{2}^{2}\overline{\xi}P_{R}}{1+c_{2}^{2}\xi P_{R}}\right)
\label{R2outc_end_b}%
\end{align}
instead of (\ref{R1_p8}) and (\ref{R2_p8}), respectively, and the union is
taken also with respect to parameter $\xi,$ with $0\leq\xi\leq1$ and
$\overline{\xi}=1-\xi.$

With variable OBRC\ bandwidth allocation, an outer bound is given as above but
with the union taken also with respect to all parameters $\eta_{MAC},\eta
_{BC}$, $i=1,2,$ such that $\eta_{MAC}+\eta_{BC}$ $=\eta$.

\emph{Proof:} Appendix F.

\subsection{Review of Capacity Results}

While, as explained above, the analysis of the IC-OBR Type-II is more complex
than that of IC-OBR Type-I, we are able to identify specific sets of
conditions on the OBRC and IC under which conclusive results can be found.
Here we review such results.

We proceed as for the IC-OBR\ Type-I and consider at first \textit{R-to-D
bottleneck conditions}, akin to (\ref{bottleneck_conditions}), which in this
case reduce to%
\begin{equation}
\eta_{MAC}\mathcal{C}(b_{1}^{2}P_{1R})\geq\eta_{BC}\mathcal{C}(c_{1}^{2}%
P_{R})\text{ and }c_{1}\geq c_{2},\label{rtd_bottleneck_2}%
\end{equation}
or symmetrically by swapping indices 1 and 2. These conditions imply that the
capacity region of the BC beween $R$ and $D_{1},D_{2}$ is completely included
in the MAC capacity region between $S_{1},S_{2}$ and $R,$ as shown in\ Fig. 5.
We show that, under appropriate channel conditions on the IC, the sum-capacity
is obtained by signal relaying only (separable operation) in Proposition 8,
similarly to Proposition 3 for IC-OBR\ Type-I.

We then consider a number of scenarios where the condition above is not
satisfied, and an excess rate between $S_{2}$ and $D_{1}$, similar to
(\ref{ex_rate_conditions}), is available on the OBRC (see Fig. 6 for an
illustration). We first show in Proposition
9 that, similarly to Proposition 4 and 5 (see Remarks 3 and 4), if decoder $D_{2}$ is already in sufficiently strong interference
conditions, then interference forwarding is not necessary. We then exhibit a set of conditions under which interference
forwarding, along with signal relaying, achieves capacity in Proposition 10.
This result is akin to Proposition 5 in that it mimics the sum-capacity result
for a Gaussian IC in the mixed-interference regime, and the excess rate on the
OBRC has the effect of driving decoder $D_{2}$ in the strong interference
regime. While the two results above are obtained via DF on the OBRC, we
finally show that CF may be optimal too in Proposition 11.

We then consider variable bandwidth allocation over the OBRC and obtain, via
numerical results, similar conclusions as for the type-I IC-OBR.

\subsection{Capacity Results for Fixed OBRC Bandwidth Allocation}

In this section, we consider fixed OBRC bandwidth allocation.

\subsubsection{R-to-D Bottleneck Regime}

The next Proposition finds the sum-capacity for the R-to-D bottleneck
regime (\ref{rtd_bottleneck_2}).

\textbf{Proposition 8 (R-to-D Bottleneck Regime):} In an IC-OBR Type-II with
fixed bandwidth allocation and (\ref{rtd_bottleneck_2}), we have that if
$a_{12}<1$, $a_{21}\geq\sqrt{\frac{1+P_{1}}{1+a_{12}^{2}P_{1}}}$, the sum
capacity is given by
\begin{equation}
R_{1}+R_{2}\leq\mathcal{C}\left(  P_{1}\right)  +\mathcal{C}\left(
\frac{P_{2}}{1+a_{12}^{2}P_{1}}\right)  +\eta_{BC}\mathcal{C}\left(  c_{1}%
^{2}P_{R}\right)  \label{Prop11ratesum1}%
\end{equation}
and is obtained by signal relaying only and separable operation.

\emph{Proof:} Appendix G.

\vspace{0.1in}

\begin{figure}[t]
\begin{center}
\includegraphics[width=10 cm]{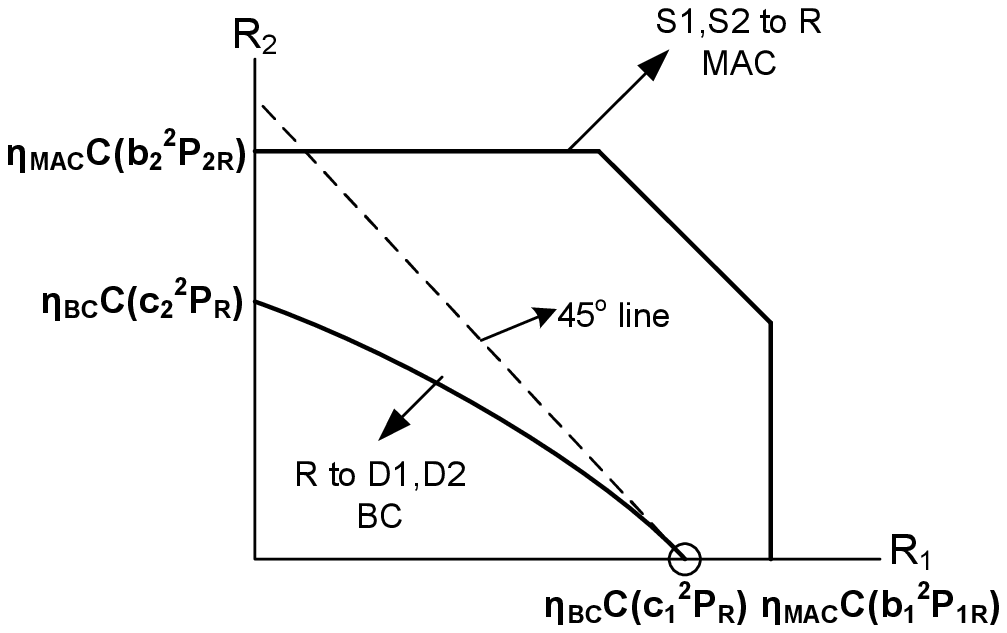}
\end{center}
\caption{Illustration of the OBRC conditions leading to the sum-capacity in
Proposition 8: $c_{1}\geq c_{2}$, $\eta_{MAC}\mathcal{C}(b_{1}^{2}P_{1R}%
)\geq\eta_{BC}\mathcal{C}(c_{1}^{2}P_{R})$.}%
\label{Fig:MAC_BC_a}%
\end{figure}

\begin{figure}[t]
\begin{center}
\includegraphics[width=10 cm]{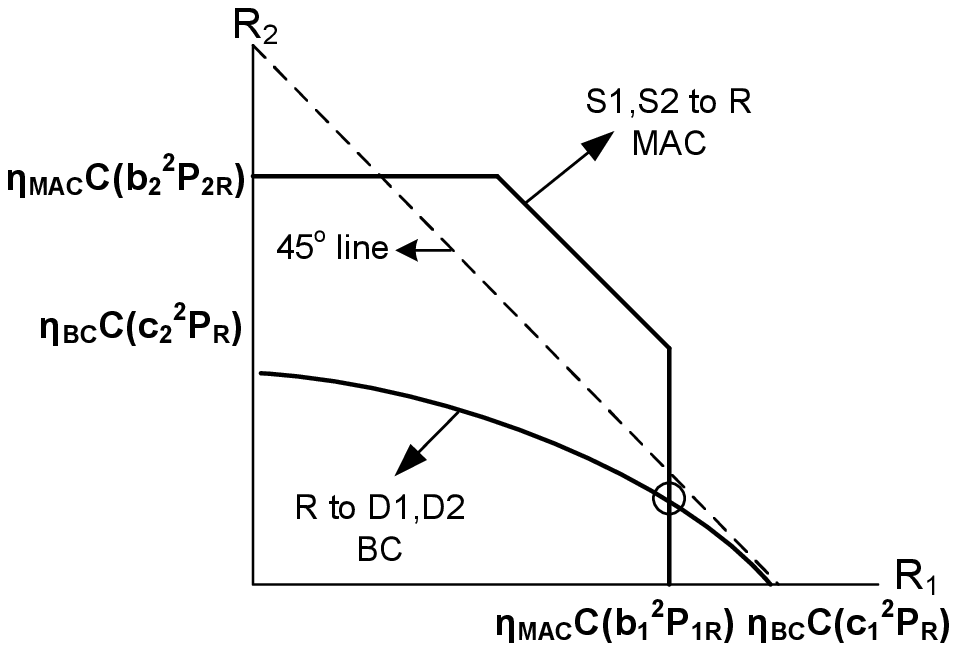}
\end{center}
\caption{Illustration of the OBRC conditions leading to the sum-capacity in
Proposition 9: $c_{1}\geq c_{2}$, $\eta_{MAC}\mathcal{C}(b_{1}^{2}P_{1R})<
\eta_{BC}\mathcal{C}(c_{1}^{2}P_{R})$, $\eta_{MAC}\mathcal{C}\left(
\frac{b_{2}^{2}P_{2R}}{1+b_{1}^{2}P_{1R}}\right) \geq\eta_{BC}C\left(
\frac{c_{2}^{2}\overline{\xi}^{\ast}P_{R}}{1+c_{2}^{2}\xi^{\ast}P_{R}}\right)
$ where where $\xi^{\ast}$ is the optimal power allocation that maximizes the
sum-rate in Proposition 12 ($\overline{\xi}^{\ast}=1-\xi^{\ast})$.}%
\label{Fig:MAC_BC_b}%
\end{figure}

\textbf{Remark 6:} In Proposition 8, the interference conditions on the IC are
\textit{mixed} and the optimal transmission strategy over the IC turns out to
prescribe transmission of only private information by $S_{1}$ (given that
$D_{2}$ is in weak interference) and of only common information by $S_{2}$
(given that $D_{1}$ is in a strong interference condition). The conditions on
the OBRC are illustrated in Fig. \ref{Fig:MAC_BC_a}. The optimal operation
over the OBRC\ in terms of sum-rate is for only user 1 to transmit over the
OBRC using signal forwarding with DF at the relay. Notice that this operating
point on the OBRC\ (see dot in Fig. \ref{Fig:MAC_BC_a}) is sum-rate optimal if
one focuses on the OBRC alone and on DF, since the corresponding achievable
rate region is given by the intersection of the MAC and BC\ regions in Fig.
\ref{Fig:MAC_BC_a}. Proposition 8 shows that such operating point is also
optimal for communications over the IC-OBR under the given conditions.\

\subsubsection{Excess Rate Regime}

We now consider the case where the R-to-D bottleneck condition
$\eta_{MAC}\mathcal{C}(b_{1}^{2}P_{1R})\geq\eta_{BC}\mathcal{C}(c_{1}^{2}%
P_{R})$ is not satisfied.

\textbf{Proposition 9 (Excess rate regime, Mixed Interference):} In an IC-OBR
Type-II with fixed bandwidth allocation, we have that if $a_{12}<1$,
$a_{21}\geq\sqrt{\frac{1+P_{1}}{1+a_{12}^{2}P_{1}}},$ then the sum-capacity is
given by
\begin{align}
R_{1}+R_{2} &  \leq\max_{0\leq\xi\leq1}\bigg\{\mathcal{C}\left(  P_{1}\right)
+\mathcal{C}\left(  \frac{P_{2}}{1+a_{12}^{2}P_{1}}\right)  +\min
\big\{\eta_{MAC}\mathcal{C}\left(  b_{1}^{2}P_{1R}\right)  ,\eta
_{BC}\mathcal{C}\left(  c_{1}^{2}\xi P_{R}\right)  \big\}\nonumber\\
&  +\eta_{BC}\mathcal{C}\left(  \frac{c_{2}^{2}\overline{\xi}P_{R}}%
{1+c_{2}^{2}\xi P_{R}}\right)  \bigg\},\label{Prop11ratesum2}%
\end{align}
if
\begin{equation}
\eta_{MAC}\mathcal{C}(b_{1}^{2}P_{1R})<\eta_{BC}\mathcal{C}(c_{1}^{2}%
P_{R})\text{ and }\eta_{MAC}\mathcal{C}\left(  \frac{b_{2}^{2}P_{2R}}%
{1+b_{1}^{2}P_{1R}}\right)  \geq\eta_{BC}C\left(  \frac{c_{2}^{2}\overline
{\xi}^{\ast}P_{R}}{1+c_{2}^{2}\xi^{\ast}P_{R}}\right)  ,\label{cond_excess}%
\end{equation}
where $\xi^{\ast}$ is the optimal power allocation that maximizes the sum-rate
(\ref{Prop11ratesum2}) with $\overline{\xi}^{\ast}=1-\xi^{\ast}$. The
sum-capacity is obtained by DF and signal relaying alone.

\emph{Proof:} Appendix H.

\textbf{Remark 7: }The conditions (\ref{cond_excess}) are illustrated in Fig.
\ref{Fig:MAC_BC_b}. Observing Fig. \ref{Fig:MAC_BC_b} and recalling
Propositions 4-5, one could guess that signal relaying is suboptimal under the
conditions of Proposition 9. In fact, these conditions entail an excess rate
between $S_{2}$ and $D_{1}$, since the maximum rate $S_{2}-R$ (i.e.,
$\eta_{MAC}\mathcal{C}(b_{2}^{2}P_{2R})$) is larger than the maximum rate
$R-D_{2}$ (i.e., $\eta_{BC}\mathcal{C}(c_{2}^{2}P_{R})$), while the opposite
is true for the path $S_{1}-R-D_{1}.$ The fact that such excess rate is not to
be exploited for interference forwarding by the optimal scheme of Proposition
9 can be interpreted in light of Remarks 3 and 4, since the condition
$a_{21}\geq\sqrt{\frac{1+P_{1}}{1+a_{12}^{2}P_{1}}},$ assumed in Proposition
9, already guarantees strong interference condition at $D_{1},$ and thus no
need for interference forwarding. We also remark that the sum-rate optimal
operation of the OBRC requires signal relaying for \textit{both} sources, not
just source 1 as above, and transmission over the OBRC at rates given by the
operating point indicated in the figure. This rate pair is characterized by
the power split $\xi^{\ast}$ in Proposition 9$.$

%\subsubsection{Excess Rate Regime}
%
%We now consider the case where the R-to-D bottleneck condition
%(\ref{rtd_bottleneck_2}) is not satisfied.

We next consider the case when a very
large excess rate between $S_{2}$ and $D_{1}$ is present, i.e.,
\begin{equation}
b_{2},c_{1}\rightarrow\infty,\label{large_excess}%
\end{equation}
and show that interference forwarding may be useful to drive $D_{1}$ in the strong
interference regime. The result is akin to Proposition 5 for IC-OBR\ Type-I.

\textbf{Proposition 10 (Excess rate regime, Mixed Interference)}: In an IC-OBR
Type-II with fixed bandwidth allocation, the following conditions give an
achievable region
\begin{subequations}
\begin{align}
R_{1} &  \leq\mathcal{C}(P_{1})+\eta_{MAC}\mathcal{C}(b_{1}^{2}P_{1R}%
)\label{Prop14R1ach}\\
R_{2} &  \leq\mathcal{C}\left(  \frac{P_{2}}{1+a_{12}^{2}P_{1}}\right)
+\eta_{BC}\mathcal{C}\left(  \frac{c_{2}^{2}\overline{\xi}P_{R}}{1+c_{2}%
^{2}\xi P_{R}}\right)  \\
R_{1}+R_{2} &  \leq\mathcal{C}(P_{1}+a_{21}^{2}P_{2})+\eta_{BC}\mathcal{C}%
(c_{1}^{2}\xi P_{R})+\eta_{BC}\mathcal{C}\left(  \frac{c_{2}^{2}\overline{\xi
}P_{R}}{1+c_{2}^{2}\xi P_{R}}\right)  \\
R_{1}+R_{2} &  \leq\mathcal{C}(P_{1}+a_{21}^{2}P_{2})+\eta_{MAC}%
\mathcal{C}(b_{1}^{2}P_{1R}+b_{2}^{2}P_{2R})\label{Prop14Rtotach}%
\end{align}
with $\xi+\overline{\xi}\leq1$ via DF and joint signal relaying and
interference forwarding. Moreover, if (\ref{large_excess}) and $a_{12}<1$, the
sum-capacity is achieved by such scheme and given by
\end{subequations}
\[
R_{1}+R_{2}\leq\mathcal{C}(P_{1})+\mathcal{C}\left(  \frac{P_{2}}{1+a_{12}%
^{2}P_{1}}\right)  +\eta_{MAC}\mathcal{C}(b_{1}^{2}P_{1R})+\eta_{BC}%
\mathcal{C}(c_{2}^{2}P_{R}).
\]

\emph{Proof:} Appendix I.

\textbf{Remark 8}: The scheme achieving (\ref{Prop14R1ach}%
)-(\ref{Prop14Rtotach}) is based on transmitting only private information over
the IC and OBRC (signal relaying) by $S_{1}$, due to the weak interference at
$D_{2},$ while $S_{2}$ transmits common information over the IC and both
common and private on the OBRC\ (joint signal relaying and interference
forwarding). This scheme is shown to be sum-rate optimal if weak interference
is seen at $D_{2}$ and a large \textquotedblleft excess rate\textquotedblright%
\ (\ref{large_excess}) is available between $S_{2}$ and $D_{1}$ so as to
essentially drive $D_{2}$ in the very strong interference regime.

\begin{figure}[t]
\begin{center}
\includegraphics[width=12 cm]{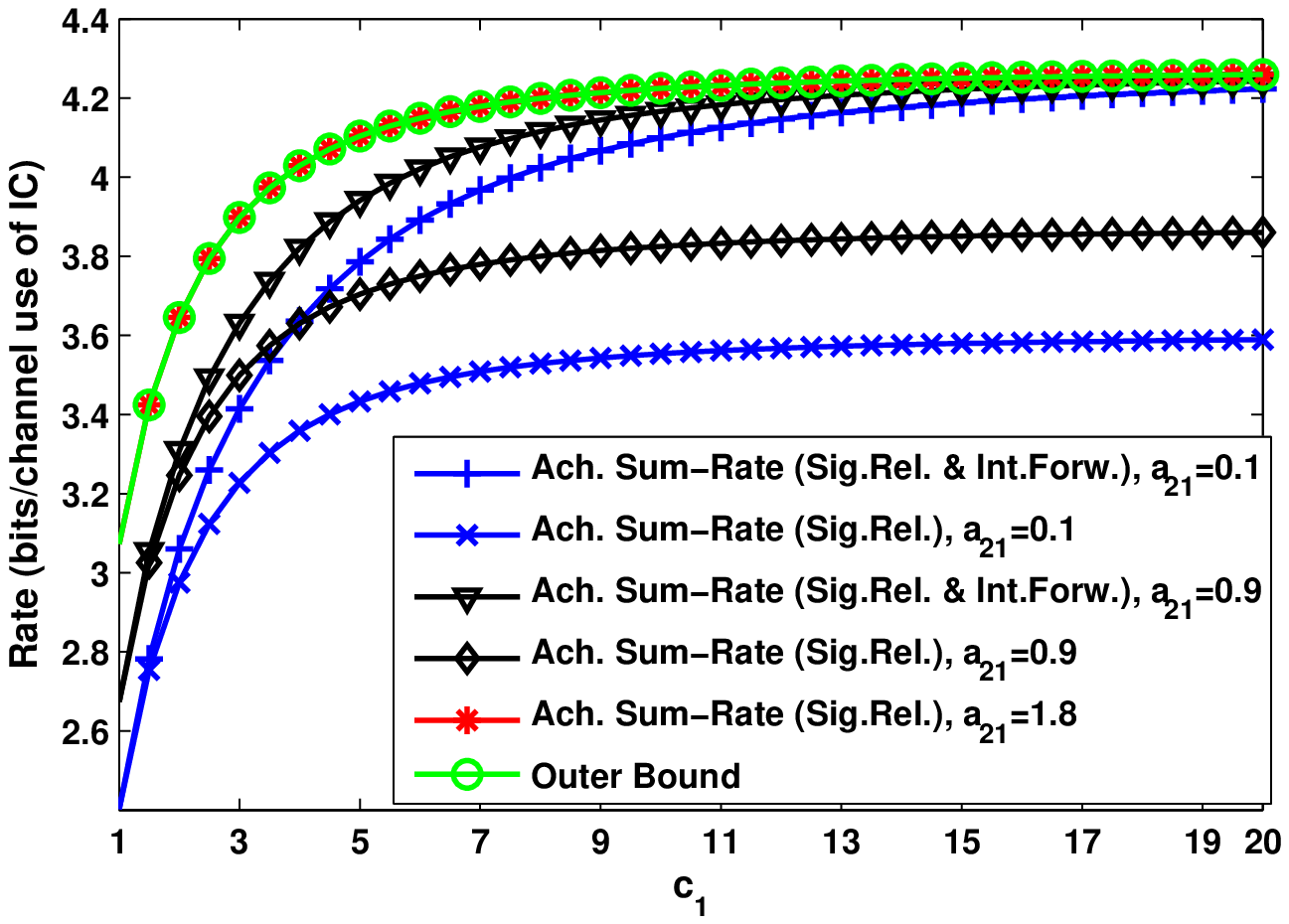}
\end{center}
\caption{Achievable sum-rate and outer bound for an IC-OBR Type-II with
respect to $R-D_{1}$ channel gain, $c_{1}$ and $S_{2}-D_{1}$ channel gain
$a_{21}\in\{0.1,0.9.1.8\}$ ($a_{12}=0.5$, $b_{1}=1$, $b_{2}=10$, $c_{2}=1$ and
all node powers are equal to 10 dB).}%
\label{Fig:ICOBRII_Int_forw}%
\end{figure}

To investigate the role of interference forwarding in a non-asymptotic regime,
Fig. \ref{Fig:ICOBRII_Int_forw} shows the sum-rate obtained from
(\ref{Prop14R1ach})-(\ref{Prop14Rtotach}), by assuming that source 2 either
uses only signal relaying (i.e., $R_{2c^{\prime}}=0$ in the achievable region
given in Appendix I) or also interference forwarding, and the sum-rate upper
bound obtained from Proposition 7 and given in (\ref{TypeIIoutbnd}), Appendix
G. The OBRC gains are set to $b_{1}=1$, $b_{2}=10$, $c_{2}=1$ and $c_{1}$ is
varied, all node powers are equal to 10 dB and $\eta_{MAC}=\eta_{BC}=1$. We
also have $a_{12}=0.5$ and $S_{2}-D_{1}$ channel gain takes the values
$a_{21}\in\{0.1,0.9.1.8\}$. Note that for $a_{21}=1.8\geq\sqrt{\frac{1+P_{1}%
}{1+a_{12}^{2}P_{1}}}=1.78$ the conditions given in Proposition 8 are
satisfied and signal relaying alone is optimal. For $a_{21}\in\{0.1,0.9\}\leq
1.78,$ the advantages of interference forwarding become substantial with
increasing $c_{1}$, which is due to the fact that the $S_{2}-D_{1}$ pair can
exploit more excess rate. The asymptotic optimality derived in Proposition 10
is here shown to be attained for finite values of $b_{2},c_{1}.$%

Finally, we briefly show that signal relaying may be (asymptotically) optimal
also in combination with CF at the relay.

\textbf{Proposition 11 (Optimality of CF): }In an IC-OBR Type-II with fixed
bandwidth allocation, the following rates are achievable via signal relaying
and CF at the relay:
\begin{subequations}
\begin{align}
R_{1} &  \leq\mathcal{C}(P_{1})+\eta_{MAC}C\left(  \frac{b_{1}^{2}P_{1R}%
}{1+\sigma^{2}}\right)  \label{cf_ach_1}\\
R_{2} &  \leq\mathcal{C}(P_{2})+\eta_{MAC}C\left(  \frac{b_{2}^{2}P_{2R}%
}{1+\sigma^{2}}\right)  \\
R_{1}+R_{2} &  \leq\min\left\{  \mathcal{C}(P_{1}+a_{21}^{2}P_{2}%
),\mathcal{C}(a_{12}^{2}P_{1}+P_{2})\right\}  +\eta_{MAC}C\left(  \frac
{b_{1}^{2}P_{1R}+b_{2}^{2}P_{2R}}{1+\sigma^{2}}\right)  ,\label{cf_ach_end}%
\end{align}
where $\sigma^{2}$ satisfies
\end{subequations}
\[
\sigma^{2}\geq\frac{1+b_{1}^{2}P_{1R}+b_{2}^{2}P_{2R}}{\min\left\{
(1+c_{1}^{2}P_{R})^{\frac{\eta_{BC}}{\eta_{MAC}}},(1+c_{2}^{2}P_{R}%
)^{\frac{\eta_{BC}}{\eta_{MAC}}}\right\}  -1}.
\]
The above provides the capacity region, given by (\ref{rtotdfob1}%
)-(\ref{CF_cap_regend}), for $a_{12}\geq1,$ $a_{21}\geq1$ and $c_{1}%
,c_{2}\rightarrow\infty.$

\textit{Proof}: Appendix J.

\begin{figure}[t]
\begin{center}
\includegraphics[width=10 cm]{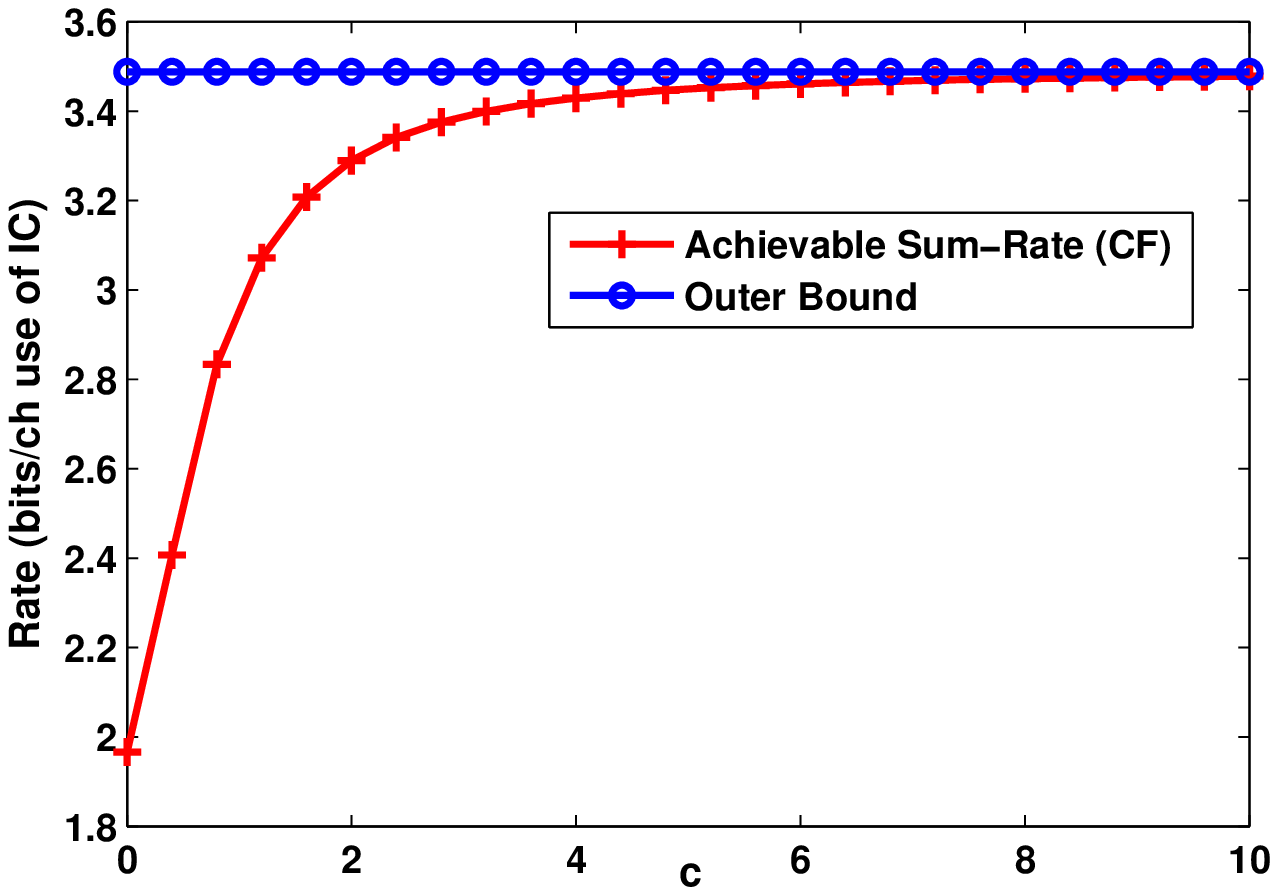}
\end{center}
\caption{Achievable sum-rate ((\ref{cf_ach_1})-(\ref{cf_ach_end})) and outer
bound ((\ref{rtotdfob1})-(\ref{CF_cap_regend})) for an IC-OBR Type-II channel
with symmetric IC and OBRC with respect to relay-to-destination channel gains,
$c$ ($a=2$, $b=1,$ $P=P_{s}=10$ dB).}%
\label{Fig:ICOBRII_CF}%
\end{figure}

%The following theorem shows that CF scheme with signal relaying is asymptotically optimal as relay to the destination channels become large.
%\textbf{Proposition 9:} In an IC-OBR, for $a_{12}\geq 1$, $a_{21}\geq 1$ and $c_1,c_2\rightarrow \infty$, the capacity region is given by,
%\begin{eqnarray}\label{CF_cap_reg1}
%R_1&\leq& \mathcal{C}( P_1 ) + \eta\mathcal{C}(b_1^2P_{1R})\\
%R_2&\leq& \mathcal{C}( P_2) + \eta\mathcal{C}(b_2^2P_{2R})\\
%R_1+R_2&\leq& \mathcal{C}( P_1 + a_{21}^2P_2 ) + \eta\mathcal{C}(b_1^2P_{1R}+b_2^2P_{2R})\\
%R_1+R_2&\leq& \mathcal{C}( a_{12}^2P_1 + P_2 ) + \eta\mathcal{C}(b_1^2P_{1R}+b_2^2P_{2R}).
%\label{CF_cap_regend}
%\end{eqnarray}
%\emph{Proof:} Appendix H.

\textbf{Remark 9:} The rate (\ref{cf_ach_1})-(\ref{cf_ach_end}) is achieved by
transmitting common information over the IC, which leads to its optimality in
the strong interference regime $a_{12}\geq1,$ $a_{21}\geq1$, and private
information (signal relaying) over the OBRC. The relay performs CF which
becomes optimal asymptotically as the relay-to-destination's BC quality
improves. Fig. \ref{Fig:ICOBRII_CF} shows the comparison of achievable
sum-rate obtained from (\ref{cf_ach_1})-(\ref{cf_ach_end}) and the outer bound
(\ref{rtotdfob1})-(\ref{CF_cap_regend}) on the sum-rate given in Proposition
11 from Appendix J for an IC-OBR Type-II channel with symmetric IC and OBRC as a function of
$c$ with $a=2$, $b=1$, $P=P_{s}=10$ dB, $\eta=1$. We observe that the achievable
sum-rate and outer bound are not only asymptotically equal for large $c,$ as
shown by Proposition 11, but in practice become very close already for
$c\geq6$.

\subsection{Capacity Results for OBRC Variable Bandwidth Allocation}

Here, we briefly investigate the effect of variable bandwidth allocation for
the IC-OBR Type-II via numerical results. We consider a mixed interference
scenario with $a_{12}=0.5$, $a_{21}=1.8$ and $c_{1}\geq c_{2}$, which
satisfies all the conditions of Proposition 8 except the ones that depend on
the bandwidth allocation ($\eta_{MAC}$, $\eta_{BC}$). Recall that for given
conditions on fixed allocations ($\eta_{MAC}$, $\eta_{BC}$), Proposition 8
shows the optimality of DF with signal relaying (separable operation). We
compare the performance of the DF scheme in Proposition 8 (separable
transmission) with an outer bound obtained from Proposition 7. In particular,
from the conditions (\ref{R1out2}) and (\ref{R1out3}) for $R_{1}$ and
(\ref{R2out1}) and (\ref{R2outc_end_b}) for $R_{2}$, with $a_{12}<1$, we
obtain the following outer bound,
\begin{align}
R_{1}+R_{2} &  \leq\mathcal{C}(P_{1})+\mathcal{C}\left(  \frac{P_{2}}%
{1+a_{12}^{2}P_{1}}\right)  +\min\left\{  \eta_{MAC}\mathcal{C}(b_{1}%
^{2}P_{1R}),\eta_{BC}\mathcal{C}(c_{1}^{2}\xi P_{R})\right\}
\nonumber\label{Rtotout_typeII_var}\\
&  +\min\left\{  \eta_{MAC}C(b_{1}^{2}P_{1R}+b_{2}^{2}P_{2R}),\eta
_{BC}C\left(  \frac{c_{2}^{2}\overline{\xi}P_{R}}{1+c_{2}^{2}\xi P_{R}%
}\right)  \right\}
\end{align}
for $\xi=1-\overline{\xi}$. In both cases, bandwidth allocation ($\eta
_{MAC},\eta_{BC})$ is optimized.

\begin{figure}[t]
\begin{center}
\includegraphics[width=9 cm]{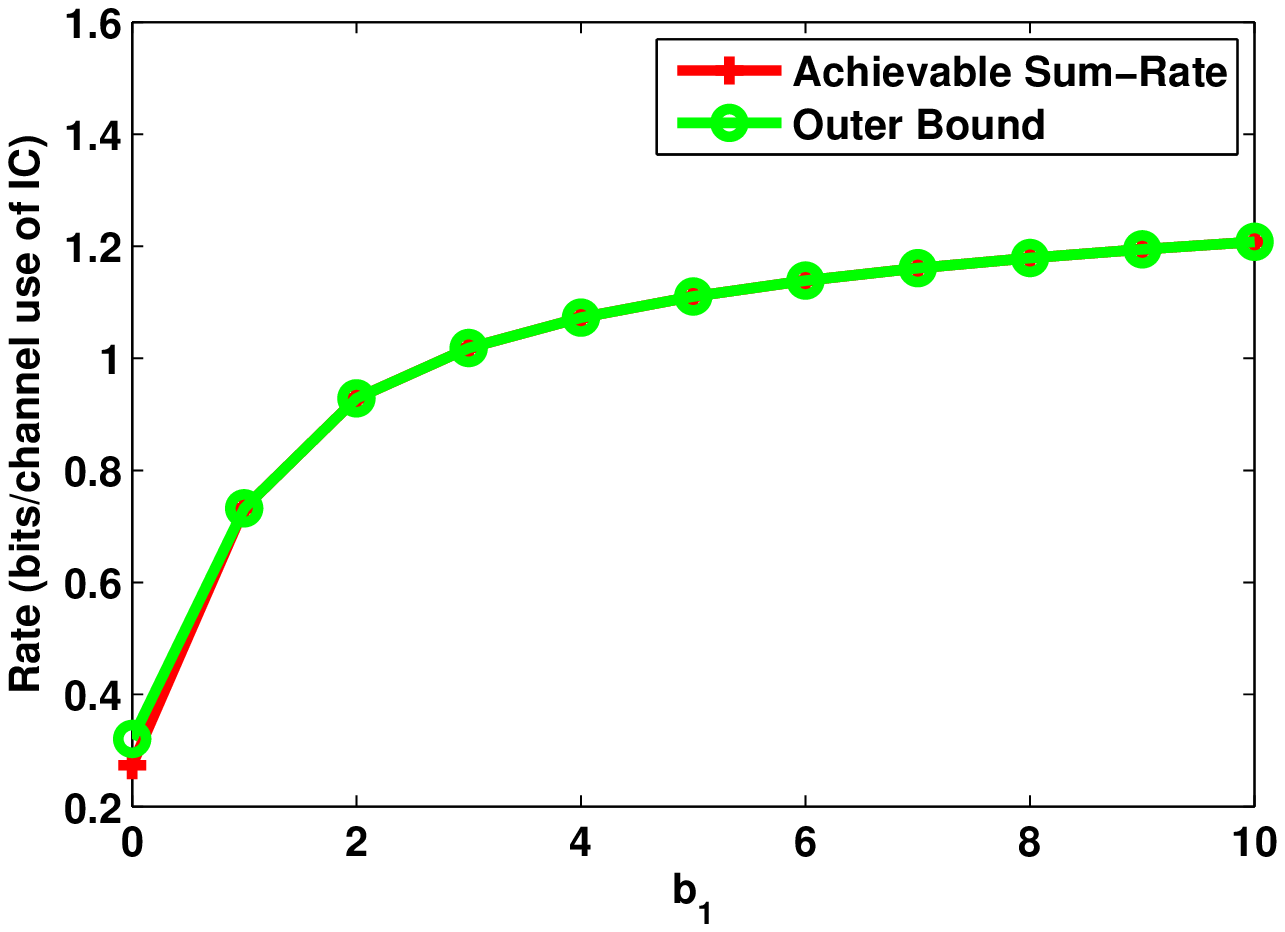}
\includegraphics[width=9 cm]{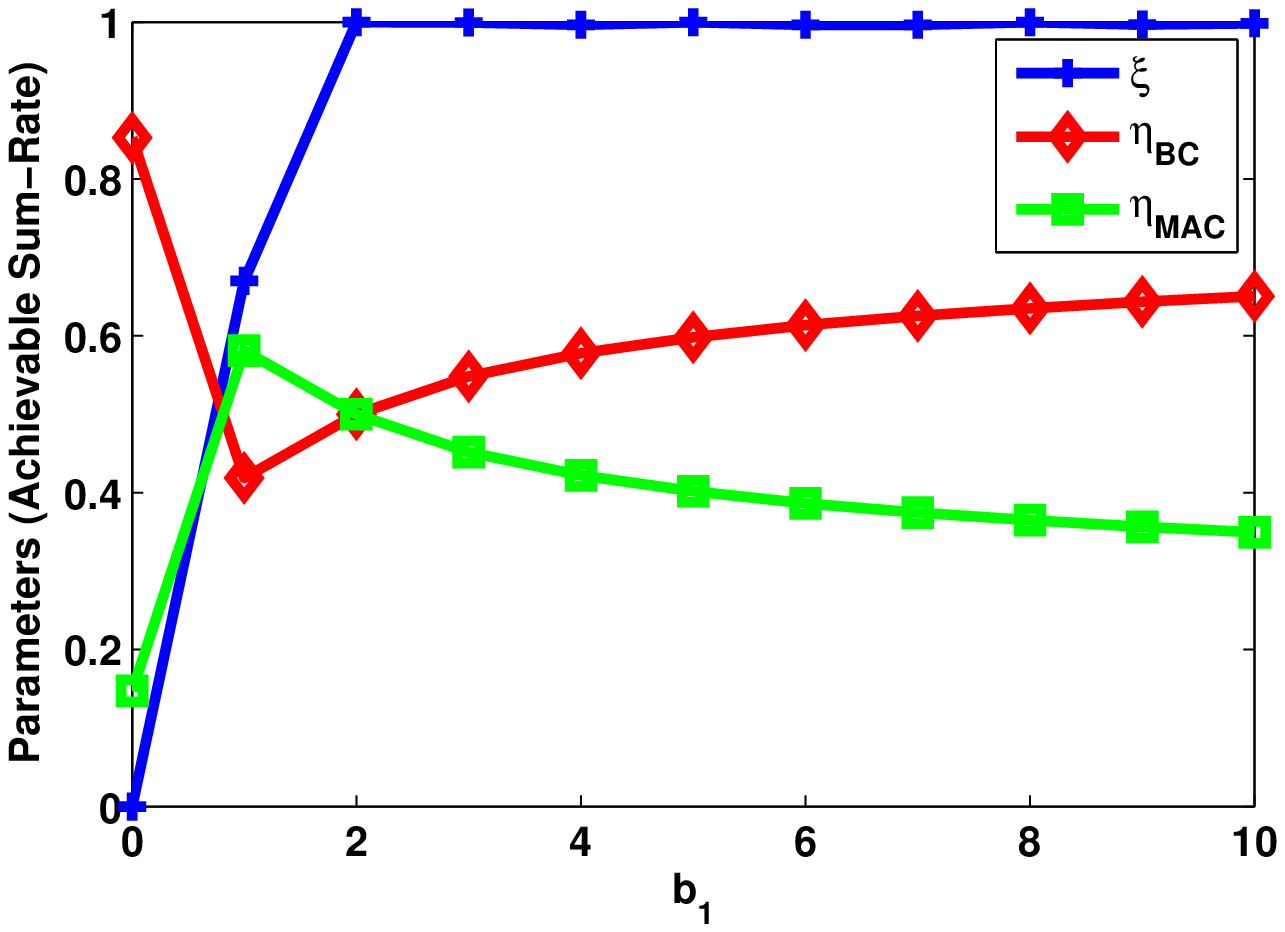}
\end{center}
\caption{ Achievable sum-rate (from Proposition 8, (\ref{Prop11ratesum1}))
with signal relaying (DF) and outer bound (\ref{Rtotout_typeII_var}) and
optimal parameters ($\eta_{MAC}$, $\eta_{BC}$, $\xi$) of the DF scheme
(\ref{Prop11ratesum1}) for an IC-OBR Type-II with respect to $S_{1}-R$ channel
gain $b_{1}$ ($b_{2}=2$, $c_{1}=2$, $c_{2}=0.3$, $\eta={1}$, all node powers
are equal to $10$ dB, $a_{21}=1.8$, $a_{12}=0.5.)$ }%
\label{Fig:Type_II_var_df_sing}%
\end{figure}

In Fig. \ref{Fig:Type_II_var_df_sing}, the sum-rate discussed above are shown
for variable $S_{1}-R$ gain, $b_{1}$, and the other channel gains are set to
$b_{2}=2$, $c_{1}=2$, $c_{2}=0.3$ and all nodes powers are $10$ dB and
$\eta=1$. The right part of the figure also shows the optimal bandwidth
($\eta_{MAC}$ and $\eta_{BC}$) and power ($\xi$) allocation. We know from
Proposition 8 that if $b_{1}$ is sufficiently larger than
$c_{1}$, for fixed bandwidth allocation, the DF rate (\ref{Prop11ratesum1})
where the relay helps the $S_{1}-D_{1}$ pair only, is optimal. A similar
conclusion is drawn here for $b_{1}\geq2$ as it can be seen from the optimal
power allocation $\xi$. Moreover, the total bandwidth is balanced between the
$S_{1}-R$ and $R-D_{1}$ channels. This result is akin to Proposition 6 for
IC-OBR Type-I.

\begin{figure}[t]
\begin{center}
\includegraphics[width=10 cm]{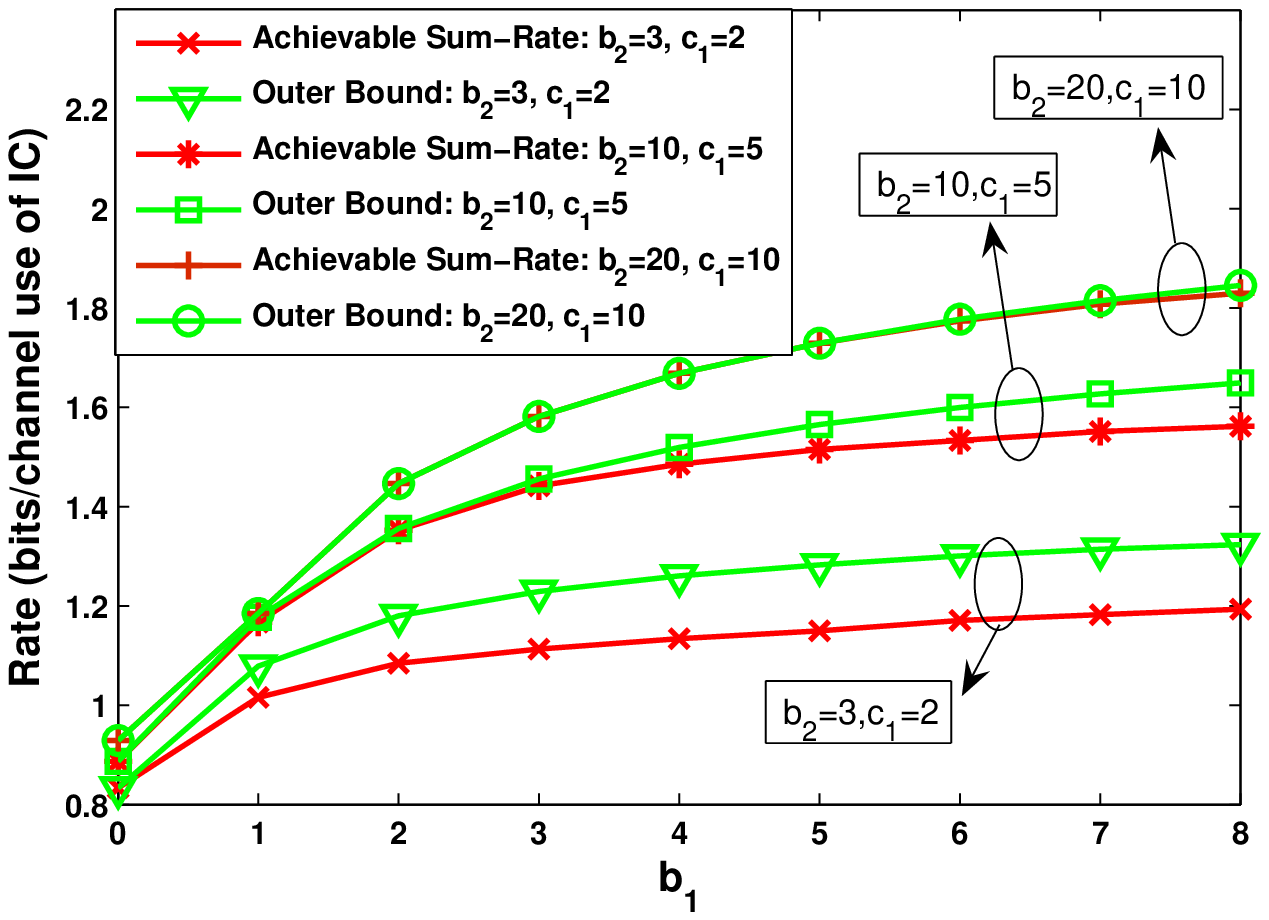}
\end{center}
\caption{ Achievable sum-rate (from Proposition 8, (\ref{Prop11ratesum1}))
with signal relaying (DF) and outer bound (\ref{Rtotout_typeII_var}) for an
IC-OBR Type-II with respect to $S_{1}-R$ channel gain, $b_{1}$ and
$(b_{2},c_{1})\in\{(3,2),(10,5),(20,10)\}$ ($c_{2}=1$, $\eta_{MAC}+\eta
_{BC}=\eta$ with $\eta={1}$, all node powers are equal to $10$ dB,
$a_{21}=1.8$, $a_{12}=0.5.)$ }%
\label{Fig:Type_II_var_dfb}%
\end{figure}

Proposition 9 proves that, for fixed bandwidth allocation, if $c_{1}$ and
$b_{2}$ are sufficiently large, it is optimal to use DF by letting the relay
help both source-destination pairs. Fig. \ref{Fig:Type_II_var_dfb} shows that
a similar conclusion holds also when optimizing the bandwidth allocation.
Specifically, Fig. \ref{Fig:Type_II_var_dfb} compares the achievable sum-rate
(\ref{Prop11ratesum2}) (attained by the DF scheme just discussed) with the
outer bound (\ref{Rtotout_typeII_var}) for variable $b_{1}$, and $b_{2}$,
$c_{1}$ with values from the set $(b_{2},c_{1})\in\{(3,2),(10,5),(20,10)\}$.
We also have $c_{2}=1$, $\eta={1}$, and other conditions as above. It is seen
that for $(b_{2},c_{1})$ large enough, the outer bound and the achievable
sum-rate match for $b_{1}\leq7$.

\begin{figure}[t]
\begin{center}
\includegraphics[width=10 cm]{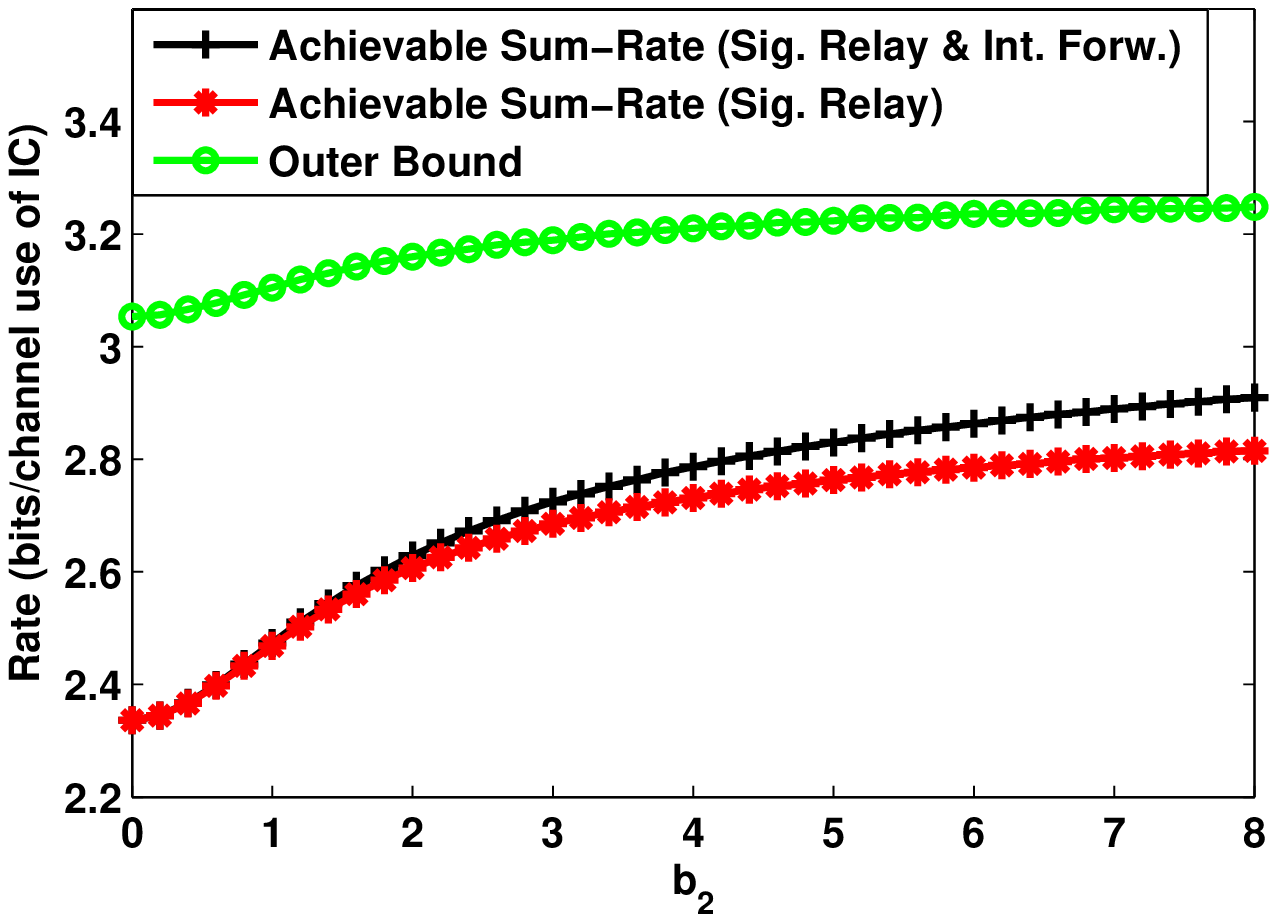}
\end{center}
\caption{Achievable sum-rate (from Proposition 10, (\ref{Prop14R1ach}%
)-(\ref{Prop14Rtotach})) with signal relaying and interference forwarding (DF,
only common information transmission by $S_{2}$ and only private information
transmission by $S_{1}$ over the IC) and outer bound (\ref{Rtotout_typeII_var}%
) versus $b_{2}$ for IC-OBR\ Type-II with variable OBRC\ bandwidth allocation
$\eta_{MAC}+\eta_{BC}=\eta,$ ($\eta=1$, $a_{21}=1$, $a_{12}=0.5$, all powers
equal to $10$ dB, $c_{1}=4$, $c_{2}=1.5$, $b_{1}=1)$.}%
\label{Fig:typeII_var_int_forw}%
\end{figure}

A natural question that arises is to understand the effect of interference
forwarding for IC-OBR Type-II with variable bandwidth allocation, similar to
its IC-OBR Type-I counterpart as discussed in Sec. \ref{Sec:TypeI_var_bw}. To
observe this effect, we consider a Type-II channel with the parameters set to
$\eta=1$, $a_{21}=1$, $a_{12}=0.5$, $c_{1}=4$, $c_{2}=1.5$, $b_{1}=1$, all
powers equal to $10$ dB and $b_{2}$ is varied. Note that since $a_{21}%
\leq\sqrt{\frac{1+P_{1}}{1+a_{12}^{2}P_{1}}}=1.78$, $D_{1}$ can favor by the
relay's interference forwarding, since the interference can not be decoded and
removed over the IC only at $D_{1}$ without affecting the sum-rate. Fig.
\ref{Fig:typeII_var_int_forw} essentially shows that this is indeed the
situation especially for $b_{2}\geq2$. In the figure, the achievable scheme
follows from Proposition 10, (\ref{Prop14R1ach})-(\ref{Prop14Rtotach}), where
$S_{1}$ transmits only private information $W_{1p}$ via the IC whereas $S_{2}$
transmits common information $(W_{2c^{\prime}},W_{2c^{\prime\prime}})$ only.
The relay facilitates interference forwarding by broadcasting $W_{2c^{\prime}%
}$ which is used at $D_{1}$ to remove part of the interference. As shown in
the figure, the increase in $b_{2}$ gain helps the OBR forward more
interference to $D_{1}$, and hence interference forwarding is crucial for
larger $b_{2}$ gains.

\section{Concluding Remarks}

\label{Sec:Conclusion}

Operation over parallel radio interfaces is bound to become increasingly
common in wireless networks due to the large number of multistandard
terminals. This enables cooperation among terminals across different
bandwidths and possibly standards. In this paper, we have studied one such
scenario where two source-destination pairs, interfering over a given
bandwidth, cooperate with a relay over an orthogonal spectral resource
(out-of-band relaying, OBR). We have focused on two different models that
correspond to distinct modes of transmission over the out-of-band relay
channel (OBRC).

As discussed in previous work, relaying can assist interfering communications
via standard signal relaying but also through interference forwarding, which
eases interference mitigation. For both considered models, this paper has
derived analytical conditions under which either signal relaying or
interference forwarding are optimal. These conditions have also been related
to the problem of assessing optimality of either separable or non-separable
transmission over parallel interference channels. Overall, the analysis shows
that, in general, joint signal relaying and interference forwarding, and thus
non-separable transmission, is necessary to attain optimal performance. This
clearly complicates the design. Moreover, this is shown to be the case for
both fixed and variable (i.e., optimized) bandwidth allocation over the OBRC.
However, in some special scenarios of interest, a separable approach has been
shown to be optimal. An example of such cases, for both considered models, is
the case where the relay has better channel conditions from the sources than
to the destination (relay-to-destinations bottleneck regime). Moreover, in the
presence of optimized OBRC bandwidth allocation, separable schemes are often
(but not always) optimal.

The analysis in this paper leaves open a number of problems related to
interference management via cooperation and through multiple radio interfaces.
In particular, scenarios that extend the current model to more than two
sources and one relay are of interest, and expected to offer new research
challenges in light of the results of \cite{cadambe_int_align}.

\section*{Appendix}

\section*{A. Proof of Proposition 1}

We consider outer bounds on $R_{1}$ and the bounds on $R_{2}$ can be obtained
similarly. We have the following bound
\begin{align}
\label{last_prop1}nR_{1}  &  =H(W_{1})\\\label{fanoprop1}
&  =I(W_{1};Y_{1}^{n},Y_{R1}^{\eta_{R1}n})+H(W_{1}|Y_{1}^{n},Y_{R1}^{\eta
_{R1}n})\\
&  \leq I(W_{1};Y_{1}^{n},Y_{R1}^{\eta_{R1}n})+n\epsilon_{n}\label{fanoprop1}%
\\
&  \leq I(X_{1}^{n};Y_{1}^{n})+h(Y_{R1}^{\eta_{R1}n})-h(Z_{R1}^{\eta_{R1}%
n})+n\epsilon_{n},\label{last_prop1end}
\end{align}
where (\ref{fanoprop1}) follows from the Fano inequality $H(W_{1}|Y_{1}%
^{n},Y_{R1}^{\eta_{R1}n})\leq n\epsilon_{n}$ and (\ref{last_prop1end}) is from
chain rule, the Markovity $W_{1}\rightarrow X_{1}^{n}\rightarrow Y_{1}^{n}$
and conditioning decreases entropy. Now, the first bound on $R_{1}$ in
(\ref{thm2new}) can be obtained by noting that, $h(Y_{R1}^{\eta_{R1}%
n})-h(Z_{R1}^{\eta_{R1}n})=h(c_{1}X_{R1}^{\eta_{R1}n}+Z_{R1}^{\eta_{R1}%
n})-h(Z_{R1}^{\eta_{R1}n})$ $\leq n\eta_{R1}\mathcal{C}(c_{1}^{2}P_{R1})$, and
therefore
\[
nR_{1}\leq I(X_{1}^{n};Y_{1}^{n})+n\eta_{R1}\mathcal{C}(c_{1}^{2}%
P_{R1})+n\epsilon_{n}.
\]
Also, from (\ref{fanoprop1}), we get the bound the other bound in
(\ref{thm2new}) from the following series of inequalities
\begin{align}
\label{R1e}nR_{1}  &  \leq I(W_{1};Y_{1}^{n},Y_{R1}^{\eta_{R1}n}%
)+n\epsilon_{n}\\
&  \leq I(W_{1};Y_{1}^{n},Y_{1R}^{\eta_{1R}n},Y_{2R}^{\eta_{2R}n}%
)+n\epsilon_{n}\label{R1b}\\\label{R1c}
&  \leq I(X_{1}^{n};Y_{1}^{n})+h(Y_{1R}^{\eta_{1R}n},Y_{2R}^{\eta_{2R}%
n})-h(Z_{1R}^{\eta_{1R}n})-h(Z_{2R}^{\eta_{2R}n})+n\epsilon_{n}\\\label{R1d}
&  \leq I(X_{1}^{n};Y_{1}^{n})+h(Y_{1R}^{\eta_{1R}n})+h(Y_{2R}^{\eta_{2R}%
n})-h(Z_{1R}^{\eta_{1R}n})-h(Z_{2R}^{\eta_{2R}n})+n\epsilon_{n}\\
&  \leq I(X_{1}^{n};Y_{1}^{n})+n\eta_{1R}\mathcal{C}(b_{1}^{2}P_{1R}%
)+n\eta_{2R}\mathcal{C}(b_{2}^{2}P_{2R})+n\epsilon_{n},
\end{align}
where (\ref{R1b}) is from the Markov chain $W_{1}\rightarrow Y_{1R}^{\eta
_{1R}n},Y_{2R}^{\eta_{2R}n}\rightarrow Y_{R1}^{\eta_{R1}n}$ and data
processing inequality, (\ref{R1c}) is from chain rule, $W_{1}\rightarrow
X_{1}^{n}\rightarrow Y_{1}^{n}$ and conditioning decreases entropy and
(\ref{R1d}) is from independence of $Y_{1R}^{\eta_{1R}n}$ and $Y_{2R}%
^{\eta_{2R}n} $. Finally, the last bound on $R_{1}$ in (\ref{thm2new}) is
obtained as,
\begin{align}
\label{R2a}nR_{1}  &  =H(W_{1}|W_{2})\\
&  =I(W_{1};Y_{1}^{n},Y_{R1}^{\eta_{R1}n}|W_{2})+H(W_{1}|Y_{1}^{n}%
,Y_{R1}^{\eta_{R1}n},W_{2})\\
&  \leq I(W_{1};Y_{1}^{n},Y_{R1}^{\eta_{R1}n}|W_{2})+n\epsilon_{n}\\\label{R2aend}
&  =I(W_{1};Y_{1}^{n}|W_{2})+I(W_{1};Y_{R1}^{\eta_{R1}n}|Y_{1}^{n}%
,W_{2})+n\epsilon_{n}\\
&  \leq I(X_{1}^{n};Y_{1}^{n}|X_{2})+h(Y_{R1}^{\eta_{R1}n}|W_{2}%
)-h(Z_{R1}^{\eta_{R1}n})+n\epsilon_{n}.
\end{align}
Now, since $h(Y_{R1}^{\eta_{R1}n}|W_{2})-h(Z_{R1}^{\eta_{R1}n})\leq
h(Y_{R1}^{\eta_{R1}n})-h(Z_{R1}^{\eta_{R1}n})\leq n\eta_{R1}\mathcal{C}%
(c_{1}^{2}P_{R1})$, we have
\[
nR_{1}\leq I(X_{1}^{n};Y_{1}^{n}|X_{2}^{n})+n\eta_{R1}\mathcal{C}(c_{1}%
^{2}P_{R1})+n\epsilon_{n}.
\]
Moreover, following (\ref{R2aend}), we have,
\begin{align}
%\label{R1c2}
nR_{1}  &  \leq I(X_{1}^{n};Y_{1}^{n}|X_{2}^{n})+I(W_{1}%
;Y_{1R}^{\eta_{1R}n},Y_{2R}^{\eta_{2R}n}|Y_{1}^{n},W_{2})+n\epsilon
_{n}\label{R1a2}\\\label{R1b2}
&  \leq I(X_{1}^{n};Y_{1}^{n}|X_{2}^{n})+h(Y_{1R}^{\eta_{1R}n},Y_{2R}%
^{\eta_{2R}n}|W_{2})-h(Z_{1R}^{\eta_{1R}n})-h(Z_{2R}^{\eta_{2R}n}%
)+n\epsilon_{n}\\\label{R1c2}
&  \leq I(X_{1}^{n};Y_{1}^{n}|X_{2}^{n})+h(Y_{1R}^{\eta_{1R}n})-h(Z_{1R}%
^{\eta_{1R}n})+n\epsilon_{n}\\
&  \leq I(X_{1}^{n};Y_{1}^{n}|X_{2}^{n})+n\eta_{1R}\mathcal{C}(b_{1}^{2}%
P_{1R})+n\epsilon_{n}%
\end{align}
where (\ref{R1a2}) is from $W_{1}\rightarrow Y_{1R}^{\eta_{1R}n},Y_{2R}%
^{\eta_{2R}n}\rightarrow Y_{R1}^{\eta_{R1}n}$, (\ref{R1b2}) is due to
conditioning decreases entropy and (\ref{R1c2}) is from the fact that
$X_{2R}^{\eta_{2R}n}$ is a function of $W_{2}$ and independence of
$Y_{1R}^{\eta_{1R}n}$ and $Z_{2R}^{\eta_{2R}n}$.

The general outer bound is then obtained by taking the union of all rates
$(R_{1},R_{2})$ satisfying the constraints for all $n\geq0$ and input
distributions $p(x^{n}_{1},x^{n}_{2})=p(x^{n}_{1})p(x^{n}_{2})$, with power
constraints $(P_{1},P_{2})$, which can be proved to coincide with the limiting
region in (\ref{thm2new}), (see, e.g., \cite{cover_mac}, Remark 1). $\Box$

\section*{B. Proof of Proposition 2}

$\emph{Codeword}$ $\emph{Generation}$ $\emph{and}$ $\emph{Encoding}$\emph{:}
The sources divide their messages as $W_{1}=(W_{1R},W_{1p},W_{1c^{\prime}%
},W_{1c^{\prime\prime}})$, and $W_{2}=(W_{2R},W_{2p},W_{2c^{\prime}%
},W_{2c^{\prime\prime}})$ as explained in Sec.\ref{Sec: Ach. Region}. Messages
$W_{ip}$ and ($W_{ic^{\prime}}$,$W_{ic^{\prime\prime}})$ are encoded into
codewords $X_{ip}^{n}$ and $X_{ic}^{n}$ with rates $R_{ip}$, $R_{ic^{\prime}%
}+R_{ic^{\prime\prime}}$ for $i=1,2$, respectively, and sent over the IC in
$n$ channel uses. Such codewords are generated i.i.d. from independent
Gaussian distributions with zero-mean and powers $P_{ip},$ $P_{ic},$
respectively. Overall, we have the transmitted codewords over the IC:
\begin{subequations}
\label{codewords}%
\begin{align}
X_{1}^{n}(W_{1})  &  =X_{1p}^{n}(W_{1p})+X_{1c}^{n}(W_{1c^{\prime}%
},W_{1c^{\prime\prime}})\\
X_{2}^{n}(W_{2})  &  =X_{2p}^{n}(W_{2p})+X_{2c}^{n}(W_{2c^{\prime}%
},W_{2c^{\prime\prime}}).
\end{align}
Message $W_{iR}$ is transmitted to $D_{i}$ via the OBRC only. Moreover, to
facilitate interference cancellation, source $S_{i}$ transmits message
$W_{ic^{^{\prime}}}$ to the interfered destination $D_{j}$, $j\neq i,$ via the
OBRC. The messages $(W_{ic^{^{\prime}}},W_{iR})$ are jointly encoded by
$S_{i}$ into the codewords $X_{iR}^{\eta_{iR}n}(W_{ic^{\prime}},W_{iR})$ which
are generated i.i.d. with rate $R_{ic^{\prime}}+R_{iR}$ from independent
Gaussian distributions with zero-mean and power $P_{iR}$, $i=1,2$. On the
other hand, after successfully decoding the messages $(W_{1R},W_{2R}%
,W_{1c^{\prime}},W_{2c^{\prime}})$, the OBR encodes these messages into the
codewords $X_{Ri}^{\eta_{Ri}n}(W_{jc^{\prime}},W_{iR})$ with rate
$R_{jc^{\prime}}+R_{iR}$ which are also generated i.i.d. from independent
Gaussian distributions with zero-mean and power $P_{Ri},$ $j\neq i$,
$i,j=1,2.$

%Thus, the messages sent over the links are given by:\
%\begin{align}
%V_{1R}  & =(W_{1c^{\prime}},W_{1R}),\text{ }V_{2R}=(W_{2c^{\prime}},W_{2R})\\
%V_{R1}  & =(W_{2c^{\prime}},W_{1R}),\text{ }V_{R2}=(W_{1c^{\prime}},W_{2R}).
%\end{align}

\emph{Decoding:} The destination $D_{i}$ initially decodes the messages
$(W_{jc^{\prime}},W_{Ri})$ using the $R-D_{i}$ channel which leads to the
achievable rates (\ref{CR1ach}) and (\ref{ach_reg_end}). The signals received
on the IC are given by (\ref{sys_mod_a}) with (\ref{codewords}). Moreover,
since the destination $D_{1}$ decodes $W_{2c^{\prime}}$, it thus sees an
equivalent codebook $X_{2c}^{n}(W_{2c^{\prime}},W_{2c^{\prime\prime}})$ with
only $2^{nR_{2c^{\prime\prime}}}$ codewords (and power $P_{ic}).$ Similarly,
$D_{2}$ sees an equivalent codebook $X_{1c}^{n}(W_{1c^{\prime}},W_{1c^{\prime
\prime}})$ with rate$\ R_{1c^{\prime\prime}}.$ Decoding of the messages
$(W_{1c^{\prime}},W_{1c^{\prime\prime}},W_{1p},W_{2c^{\prime\prime}})$ at
destination $D_{1}$ (and $(W_{2c^{\prime}},W_{2c^{\prime\prime}}%
,W_{2p},W_{1c^{\prime\prime}})$ at destination $D_{2})$ is then performed
jointly as over a multiple access channel with three sources of rates
$R_{1c}=R_{1c^{\prime}}+R_{1c^{\prime\prime}},$ $R_{1p}$ and $R_{2c^{\prime
\prime}}$ (and $R_{2c}=R_{2c^{\prime}}+R_{2c^{\prime\prime}},$ $R_{2p}$ and
$R_{1c^{\prime\prime}}$ for $D_{2}$)$,$ by treating the private messages as
noise, thus with equivalent noise power $N_{i}=a_{ji}^{2}P_{jp}+1$ for
$i,j=1,2$, $i\neq j$ hence giving the achievable rates (\ref{CS1}) and
(\ref{CS2}). It is also noted that, as explained in \cite{Motani}, error
events corresponding to erroneous decoding of only message $W_{2c^{\prime
\prime}}$ at destination $D_{1}$ and $W_{1c^{\prime\prime}}$ at destination
$D_{2}$ do not contribute to the probability of error and thus can be
neglected. The relay decodes the messages $(W_{ic^{\prime}},W_{iR})$, $i=1,2$
using the orthogonal source-to-relay links as given in (\ref{YiR}). Therefore,
it is possible to show that the rates in (\ref{C1Rach}) and (\ref{C2Rach})
which are the point-to-point rates in decoding the messages $(W_{iR}%
,W_{ic^{\prime}})$ are achievable. $\Box$

\section*{C. Proof of Proposition 4}

We start with part (\textit{i}). The converse follows from Proposition 1.
Namely, the upper bounds on individual rates (\ref{R1th4}) and (\ref{R2th4})
are a consequence of the second bounds on both $R_{1}$ and $R_{2}$, while the
upper bound on the sum rate (\ref{Rtotth4}) follows by summing second and
first bounds on $R_{1}$, $R_{2}$, respectively and accounting for the
condition $a_{12}\geq1$ as
\end{subequations}
\begin{align}
R_{1}+R_{2} &  \leq\frac{1}{n}I(X_{1}^{n};Y_{1}^{n}|X_{2}^{n})+\frac{1}%
{n}I(X_{2}^{n};Y_{2}^{n})+\min\left(  \eta_{1R}\mathcal{C}(b_{1}^{2}%
P_{1R}),\eta_{R1}\mathcal{C}(c_{1}^{2}P_{R1}\right)
)\nonumber\\
&  +\min\left(  \eta_{1R}\mathcal{C}(b_{1}^{2}P_{1R})+\eta_{2R}\mathcal{C}%
(b_{2}^{2}P_{2R}),\eta_{R2}\mathcal{C}(c_{2}^{2}P_{R2})\right)  \\
&  =\frac{1}{n}h(X_{1}^{n}+Z_{1}^{n})-h(Z_{1}^{n})+\frac{1}{n}h(a_{12}%
X_{1}^{n}+X_{2}^{n}+Z_{2}^{n})-\frac{1}{n}h(a_{12}X_{1}^{n}+Z_{2}%
^{n})\nonumber\\
&  +\eta_{1R}\mathcal{C}(b_{1}^{2}P_{1R})+\eta_{R2}\mathcal{C}(c_{2}^{2}%
P_{R2})\label{Rtotprfafin}\\
&  \leq\mathcal{C}(a_{12}^{2}P_{1}+P_{2})+\eta_{1R}\mathcal{C}(b_{1}^{2}%
P_{1R})+\eta_{R2}\mathcal{C}(c_{2}^{2}P_{R2})\label{Rtotprfdfin}
\end{align}
where (\ref{Rtotprfafin}) is due to the conditions $\eta_{1R}\mathcal{C}%
(b_{1}^{2}P_{1R})\leq\eta_{R1}\mathcal{C}(c_{1}^{2}P_{R1})$ and $\eta
_{R2}\mathcal{C}(c_{2}^{2}P_{R2})\leq\eta_{2R}\mathcal{C}(b_{2}^{2}P_{2R})$,
(\ref{Rtotprfdfin}) is from the worst-case noise result \cite{worst-case}, i.e.,
$h(X_{2}^{n}+Z_{2}^{n})-h(a_{21}X_{2}^{n}+Z_{1}^{n})\leq n\log(1)$ for
$a_{21}\geq1$, and the first entropy is maximized by i.i.d. Gaussian inputs.

For achievability, we use the general result of Proposition 2, where the
sources transmit common messages $(W_{1c^{\prime\prime}},W_{2c^{\prime\prime}%
})$ over the IC which are decoded at both destinations. In addition, $S_{2}$
transmits also the message $W_{2c^{\prime}}$ to be decoded at $D_{2}$. The
other rates are set to $R_{1c^{\prime}}=R_{1p}=R_{2p}=0.$ The OBRC is used to
transmit independent messages $W_{1R},W_{2R}$ with rates $R_{1R}=\eta
_{1R}\mathcal{C}(b_{1}^{2}P_{1R})$ and $R_{2R}=\eta_{R2}\mathcal{C}(c_{2}%
^{2}P_{R2})$, but also message $W_{2c^{\prime}}$ of rate $R_{2c^{\prime}}$ to
$D_{1}$ in order to facilitate interference cancellation$.$ From Proposition
2, and applying Fourier-Motzkin elimination, we obtain the following
achievable region
\begin{align}
R_{1} &  \leq\mathcal{C}(P_{1})+\eta_{1R}\mathcal{C}(b_{1}^{2}P_{1R}%
)\label{ach_reg_thm4}\\
R_{2} &  \leq\mathcal{C}(P_{2})+\eta_{R2}\mathcal{C}(c_{2}^{2}P_{R2})\\
R_{1}+R_{2} &  \leq\mathcal{C}(a_{12}^{2}P_{1}+P_{2})+\eta_{1R}\mathcal{C}%
(b_{1}^{2}P_{1R})+\eta_{R2}\mathcal{C}(c_{2}^{2}P_{R2})\label{thm4_1}\\
R_{1}+R_{2} &  \leq\mathcal{C}(P_{1}+a_{21}^{2}P_{2})+R_{ex_{21}}+\eta
_{1R}\mathcal{C}(b_{1}^{2}P_{1R})+\eta_{R2}\mathcal{C}(c_{2}^{2}%
P_{R2}),\label{thm4_2}%
\end{align}
so that for $R_{ex_{21}}$ $\geq$ $\max\{0,\mathcal{C}(a_{12}^{2}P_{1}%
+P_{2})-\mathcal{C}(P_{1}+a_{21}^{2}P_{2})\}$, the claim is proved.

We now move to part (\textit{ii}). The converse is again a consequence of
Proposition 1. Specifically, the single rate bounds (\ref{R1th5}) and
(\ref{R2th5}) follow immediately from the second bounds on $R_{1}$ and $R_{2}%
$, while the bound on the sum-rate (\ref{Rtotth5}) is obtained from the
summation of first and second bound on $R_{1}$, $R_{2}$, respectively, and the
condition $a_{12}\geq1$ as \vspace{-0.05in}%
\begin{align}
R_{1}+R_{2} &  \leq\frac{1}{n}I(X_{1}^{n};Y_{1}^{n})+\frac{1}{n}I(X_{2}%
^{n};Y_{2}^{n}|X_{1}^{n})+\min(\eta_{1R}\mathcal{C}(b_{1}^{2}P_{1R})+\eta
_{2R}\mathcal{C}(b_{2}^{2}P_{2R}),\eta_{R1}\mathcal{C}(c_{1}^{2}%
P_{R1}))\nonumber\label{Rtot5prfc}\\
&  +\min(\eta_{2R}\mathcal{C}(b_{2}^{2}P_{2R}),\eta_{R2}\mathcal{C}(c_{2}%
^{2}P_{R2}))\\\label{Rtot5prfb}
&  =\frac{1}{n}h(X_{1}^{n}+a_{21}X_{2}^{n}+Z_{1}^{n})-\frac{1}{n}h(a_{21}%
X_{2}^{n}+Z_{1}^{n})+\frac{1}{n}h(X_{2}^{n}+Z_{2}^{n})-h(Z_{2}^{n})\nonumber\\
&  +\eta_{R1}\mathcal{C}(c_{1}^{2}P_{R1})+\eta_{R2}\mathcal{C}(c_{2}^{2}%
P_{R2})\\
&  \leq\mathcal{C}(P_{1}+a_{21}^{2}P_{2})+\eta_{R2}\mathcal{C}(c_{2}^{2}%
P_{R2})+\eta_{R1}\mathcal{C}(c_{1}^{2}P_{R1}),\label{Rtot5prfefin}
\end{align}
where (\ref{Rtot5prfb}) is due to the conditions $\eta_{R1}\mathcal{C}%
(c_{1}^{2}P_{R1})+\eta_{R2}\mathcal{C}(c_{2}^{2}P_{R2})\leq\eta_{1R}%
\mathcal{C}(b_{1}^{2}P_{1R})+\eta_{2R}\mathcal{C}(b_{2}^{2}P_{2R})$ and
$\eta_{R2}\mathcal{C}(c_{2}^{2}P_{R2})\leq\eta_{2R}\mathcal{C}(b_{2}^{2}%
P_{2R})$, (\ref{Rtot5prfefin}) is from the worst-case noise result
\cite{worst-case}, i.e., $h(X_{2}^{n}+Z_{2}^{n})-h(a_{21}X_{2}^{n}+Z_{1}%
^{n})\leq n\frac{1}{2}\log(1)$ for $a_{21}\geq1$, and the fact that first
entropy is maximized by i.i.d. Gaussian inputs.

For the achievability, consider the achievable rate region given in the proof
of Proposition 4 ((\ref{ach_reg_thm4})-(\ref{thm4_2})) (\ref{R1th4}%
)-(\ref{Rtotth4}). Clearly, when the conditions in Proposition 5 which can
also be written as $\eta_{R1} \mathcal{C}(c_{1}^{2}P_{R1})-\eta_{1R}%
\mathcal{C}(b_{1}^{2}P_{1R})\leq\eta_{2R}\mathcal{C}(b_{2}^{2}P_{2R}%
)-\eta_{R2} \mathcal{C}(c_{2}^{2}P_{R2})$ and $R_{ex_{21}}\leq\mathcal{C}%
(a_{12}^{2}P_{1}+P_{2})-\mathcal{C}(P_{1}a_{21}^{2}P_{2})$ are satisfied,
(\ref{Rtot5prfefin}) is achievable, hence gives the sum capacity. $\Box$

\section*{D. Proof of Proposition 5}

The converse is obtained from Proposition 1 by adding the second constraint on
$R_{1}$ and first constraint on $R_{2}$ in (\ref{thm2new}) such that,
\begin{align}
R_{1}+R_{2}  &  \leq\frac{1}{n}I(X_{1}^{n};Y_{1}^{n}%
|X_{2}^{n})+ \frac{1}{n}I(X_{2}^{n};Y_{2}^{n}) + \min(\eta_{1R}\mathcal{C}%
(b_{1}^{2}P_{1R}),\eta_{R1}\mathcal{C}(c_{1}^{2}P_{R1}))\nonumber\\
&  + \min( \eta_{1R}\mathcal{C}(b_{1}^{2}P_{1R})+\eta_{2R}\mathcal{C}%
(b_{2}^{2}P_{2R}),\eta_{R2}\mathcal{C}(c_{2}^{2}P_{R2}))\\
&  =\frac{1}{n}h(X_{1}^{n} + Z_{1}^{n})-\frac{1}{n}h(Z_{1}^{n})+\frac{1}%
{n}h(a_{12}X_{1}^{n} + X_{2}^{n} + Z_{2}^{n})-\frac{1}{n}h(a_{12}X_{1}^{n} +
Z_{2}^{n})\nonumber\\
& +\eta_{1R}\mathcal{C}(b_{1}^{2}P_{1R})+\eta_{R2}\mathcal{C}(c_{2}^{2}%
P_{R2})\label{prop6stp1}\\\label{prop6stp2}
&  \leq\mathcal{C}(P_{1})+ \mathcal{C}\left(  \frac{P_{2}}{1+a_{12}^{2}P_{1}%
}\right)  + \eta_{1R}\mathcal{C}(b_{1}^{2}P_{1R}) +\eta_{R2}\mathcal{C}%
(c_{2}^{2}P_{R2})
\end{align}
where (\ref{prop6stp1}) is due to the conditions $\eta_{R1}\mathcal{C}%
(c_{1}^{2}P_{R1})\geq\eta_{1R}\mathcal{C}(b_{1}^{2}P_{1R})$ and $\eta
_{2R}\mathcal{C}(b_{2}^{2}P_{2R})\geq\eta_{R2}\mathcal{C}(c_{2}^{2}P_{R2})$,
(\ref{prop6stp2}) is from the worst-case noise result of \cite{worst-case},
i.e. $h(X_{1}^{n}+Z_{1}^{n})-h(a_{12}X_{1}^{n}+Z_{2}^{n})\leq n\mathcal{C}%
(P_{1})-n\mathcal{C}(a_{12}^{2}P_{1})$ for $a_{12}<1$.

Achievability follows directly from Proposition 2 by letting transmitter
$S_{1}$ transmit private message only, i.e., $W_{1p}$ over the IC and $W_{1R}$
over the OBR, whereas user $S_{2}$ transmits common information both on the IC
and OBR $(W_{2c^{\prime\prime}},W_{2c^{\prime}})$ as well as private message
via OBR, $W_{2R}$. The other rates are set to zero $R_{1c^{\prime}%
}=R_{1c^{\prime\prime}}=R_{2p}=0$. Then, using Fourier-Motzkin elimination, it
is possible to show that the following sum-rate is achievable,
\begin{align}
R_{1}+R_{2}  &  \leq\mathcal{C}(P_{1}+a_{21}^{2}P_{2})+R_{ex_{21}}+\eta
_{1R}\mathcal{C}(b_{1}^{2}P_{1R})+\eta_{R2}\mathcal{C}(c_{2}^{2}%
P_{R2})\label{prop6acha}\\
R_{1}+R_{2}  &  \leq\mathcal{C}(P_{1}) + \mathcal{C}\left(  \frac{P_{2}%
}{1+a_{12}^{2}P_{1}}\right)  +\eta_{1R}\mathcal{C}(b_{1}^{2}P_{1R})+\eta
_{R2}\mathcal{C}(c_{2}^{2}P_{R2})\label{prop6achb}%
\end{align}
where $R_{ex_{21}}$ is given in (\ref{ex_rate}). Then, for $R_{ex_{21}}%
\geq\mathcal{C}(P_{1}) + \mathcal{C}\left(  \frac{P_{2}}{1+a_{12}^{2}P_{1}%
}\right)  -\mathcal{C}\left(  P_{1} + a_{21}^{2}P_{2}\right)  $,
(\ref{prop6achb}) is achievable, hence gives the sum capacity.$\Box$

\section*{E. Proof of Proposition 6}

For the achievable scheme, we consider a special case of Proposition 2, where
sources operate separately over the IC and OBRC, by sending only message
$(W_{1R},W_{2R})$ over the OBRC\ (signal relaying) and only common information
$(W_{1c^{\prime\prime}},W_{2c^{\prime\prime}})$ over the IC. From Proposition
2, i.e. using (\ref{CS1})-(\ref{ach_reg_end}), we obtain that the following
sum-rate is achievable
\begin{align}
R_{1}+R_{2}  &  \leq\mathcal{C}\left(  a_{12}^{2}P_{1}+P_{2}\right)
+\min\left\{  \eta_{1R}\mathcal{C}(b_{1}^{2}P_{1R}),\eta_{R1}\mathcal{C}%
(c_{1}^{2}P_{R1}))\right\} \nonumber\label{ach_rate_sum_var}\\
&  +\min\left\{  \eta_{2R}\mathcal{C}(b_{2}^{2}P_{2R}),\eta_{R2}%
\mathcal{C}(c_{2}^{2}P_{R2}))\right\}
\end{align}
to be maximized over $(\eta_{iR},\eta_{Ri}) $\ with constraint $\sum_{i=1}%
^{2}(\eta_{iR}+\eta_{Ri})=\eta$. Optimizing over the bandwidth allocation, and
recalling that $\mathcal{C}(b_{1}^{2}P_{1R})\geq\mathcal{C}(b_{2}^{2}P_{2R})$
and $\mathcal{C}(c_{1}^{2}P_{R1})\geq\mathcal{C}(c_{2}^{2}P_{R2})$, the
optimal allocations are $\eta_{1R}^{\ast}=\eta_{R1}^{\ast}=\eta/2,$ so that
the optimal achievable sum-rate is%
\[
R_{1}+R_{2}\leq\mathcal{C}\left(  a_{12}^{2}P_{1}+P_{2}\right)  +\frac
{\eta\mathcal{C}(c_{1}^{2}P_{R1})}{2}.
\]
For the outer bound, using Proposition 1 with $a_{12}=a_{21}\geq1$,
$P_{1}=P_{2}$, from (\ref{thm2new}), we obtain the upper bounds
\begin{align}
R_{1}+R_{2}  &  \leq\mathcal{C}\left(  a_{12}^{2}P_{1}+P_{2}\right)
+\min\{\eta_{1R}\mathcal{C}(b_{1}^{2}P_{1R})+\eta_{2R}\mathcal{C}(b_{2}%
^{2}P_{2R}),\eta_{R2}\mathcal{C}(c_{2}^{2}P_{R2})\}\nonumber\label{TypeI_OB}\\
&  +\min\{\eta_{1R}\mathcal{C}(b_{1}^{2}P_{1R}),\eta_{R1}\mathcal{C}(c_{1}%
^{2}P_{R1})\}
\end{align}
which should be maximized over $(\eta_{iR},\eta_{Ri})$ with $\sum_{i=1}%
^{2}(\eta_{iR}+\eta_{Ri})=\eta$. Optimizing over $(\eta_{iR},\eta_{Ri}),$
$i=1,2,$ using $\mathcal{C}(b_{1}^{2}P_{1R})\geq\mathcal{C}(b_{2}^{2}P_{2R})$,
the optimal allocations satisfy, $\eta_{2R}^{*}=0$, $\eta_{1R}^{*}+\eta
_{R1}^{*}+\eta_{R2}^{*}=\eta$, hence the optimization problem becomes,
\begin{align}
R_{1}+R_{2}  &  \leq\mathcal{C}\left(  a_{12}^{2}P_{1}+P_{2}\right)
+\min\{\eta_{1R}\mathcal{C}(b_{1} ^{2}P_{1R}),\eta_{R2}\mathcal{C}(c_{2}%
^{2}P_{R2})\}\nonumber\\
&  +\min\{\eta_{1R}\mathcal{C}(b_{1}^{2}P_{1R}),\eta_{R1}\mathcal{C}(c_{1}%
^{2}P_{R1})\}.
\end{align}
Since both $\min$ terms are limited by the same expression, $\eta
_{1R}\mathcal{C}(b_{1}^{2}P_{1R})$, the optimal bandwidth allocation will lead
to the largest among these two terms. On the other hand, optimizing the $\min$
terms individually, we have,
\begin{align}
R_{1}+R_{2}  &  \leq\max\bigg\{\mathcal{C}\left(  a_{12}^{2}P_{1}%
+P_{2}\right)  +\frac{2\eta\mathcal{C}(c_{1}^{2}P_{R1})\mathcal{C}(c_{2}%
^{2}P_{R2})}{\mathcal{C}(c_{1}^{2}P_{R1})+2\mathcal{C}(c_{2}^{2}P_{R2}%
)},\nonumber\\
&  \mathcal{C}\left(  a_{12}^{2}P_{1}+P_{2}\right)  +\frac{\eta\mathcal{C}%
(c_{1}^{2}P_{R1})}{2}\bigg\},\label{rate_sum_var}%
\end{align}
where the first term in the $\max$ corresponds to the choice $\eta_{R2}^{\ast}
=\eta_{1R}^{\ast}\frac{\mathcal{C}(b_{1}^{2}P_{1R})}{\mathcal{C}(c_{2}%
^{2}P_{R2})},\mbox{  }\eta_{R1}^{\ast}=\eta_{1R}^{\ast},\mbox{  }\eta
_{1R}^{\ast}+\eta_{R1}^{\ast}+\eta_{R2}^{\ast}=\eta,$ whereas the second to
$\eta_{1R}^{\ast}=\eta_{R1}^{\ast}=\frac{\eta}{2},\eta_{R2}^{\ast}=0$. It is
possible to show that for
\[
\mathcal{C}(c_{1}^{2}P_{R1})\geq2\mathcal{C}(c_{2}^{2}P_{R2})
\]
the outer bound obtained in (\ref{rate_sum_var}) becomes equal to the optimal
achievable sum-rate.$\Box$

\section*{F. Proof of Proposition 7}

We start with the bound (\ref{R1out1}):%
\begin{align}
nR_{1} &  \leq H(W_{1})\\
&  \leq I(W_{1};Y_{1}^{n},Y_{R1}^{\eta_{BC}n})+H(W_{1}|Y_{1}^{n},Y_{R1}%
^{\eta_{BC}n})\\\label{Fanonend}
&  \leq I(W_{1};Y_{1}^{n},Y_{R1}^{\eta_{BC}n})+n\epsilon_{n}\\\label{R2mod}
&  =I(W_{1};Y_{1}^{n})+I(W_{1};Y_{R1}^{\eta_{BC}n}|Y_{1}^{n})+n\epsilon_{n}\\\label{Markov1}
&  \leq I(X_{1}^{n};Y_{1}^{n})+I(W_{1};Y_{R}^{\eta_{MAC}n}|Y_{1}%
^{n})+n\epsilon_{n}\\
&  \leq I(X_{1}^{n};Y_{1}^{n})+h(Y_{R}^{\eta_{MAC}n})-h(Z_{R}^{\eta_{MAC}%
n})+n\epsilon_{n}\\
&  \leq I(X_{1}^{n};Y_{1}^{n})+h(b_{1}X_{1R}^{\eta_{MAC}n}+b_{2}X_{2R}%
^{\eta_{MAC}n}+Z_{R}^{\eta_{MAC}n})-h(Z_{R}^{\eta_{MAC}n})+n\epsilon_{n}\\
&  \leq I(X_{1}^{n};Y_{1}^{n})+\eta_{MAC}n\mathcal{C}\left(  b_{1}^{2}%
P_{1R}+b_{2}^{2}P_{2R}\right)  +n\epsilon_{n}%
\end{align}
where (\ref{Fanonend}) is from Fano's inequality, (\ref{Markov1}) is from the
Markov relations $W_{1}\rightarrow Y_{R}^{\eta_{MAC}n},Y_{1}^{n}\rightarrow
Y_{R1}^{\eta_{BC}n}$ and $W_{1}\rightarrow X_{1}^{n}\rightarrow Y_{1}^{n}$.

From cut-set bound around $S_{1}$, we obtain the bound (\ref{R1out2}) as
\begin{align}
nR_{1}  &  \leq I(X_{1}^{n};Y_{1}^{n}|X_{2}^{n})+I(X_{1R}^{\eta_{MAC}n}%
;Y_{R}^{\eta_{MAC}n}|X_{2R}^{\eta_{MAC}n})\\
&  \leq I(X_{1}^{n};Y_{1}^{n}|X_{2}^{n})+\eta_{MAC}n\mathcal{C}\left(
b_{1}^{2}P_{1R}\right)  .
\end{align}
Using similar steps we obtain the corresponding bounds on $R_{2}$
(\ref{R2out1}) and (\ref{R2out2})\ as
\begin{align}
nR_{2}  &  \leq I(X_{2}^{n};Y_{2}^{n})+\eta_{MAC}n\mathcal{C}\left(  b_{1}%
^{2}P_{1R}+b_{2}^{2}P_{2R}\right) \\
nR_{2}  &  \leq I(X_{2}^{n};Y_{2}^{n}|X_{1}^{n})+\eta_{MAC}n\mathcal{C}\left(
b_{2}^{2}P_{2R}\right)  .
\end{align}

We now focus on the bounds (\ref{R1out3}) and (\ref{R2outc_end_b}). For
(\ref{R2outc_end_b}), we have (from (\ref{R2mod}) modified for $R_{2}$),
\begin{align}
nR_{2} &  \leq I(X_{2}^{n};Y_{2}^{n})+I(W_{2};Y_{R2}^{\eta_{BC}n}|Y_{2}%
^{n})+n\epsilon_{n}\label{cond_dec_a}\\
&  \leq I(X_{2}^{n};Y_{2}^{n})+h(Y_{R2}^{\eta_{BC}n})-h(c_{2}X_{R}^{\eta
_{BC}n}+Z_{R2}^{\eta_{BC}n}|Y_{2}^{n},W_{2})+n\epsilon_{n}\label{R2stepend}%
\end{align}
where (\ref{R2stepend}) is from conditioning decreases entropy. Now, consider
the following,
\[
h(Z_{R2}^{\eta_{BC}n})\leq h(c_{2}X_{R}^{\eta_{BC}n}+Z_{R2}^{\eta_{BC}n}%
|Y_{2}^{n},W_{2})\leq h(c_{2}X_{R}^{\eta_{BC}n}+Z_{R2}^{\eta_{BC}n})\leq
\frac{\eta_{BC}n}{2}\log(2\pi e(1+c_{2}^{2}P_{R})).
\]
Hence, without loss of generality, one can assume
\begin{align}
h(Y_{R2}^{\eta_{BC}n}|Y_{2}^{n},W_{2})=\frac{\eta_{BC}n}{2}\log(2\pi
e(1+c_{2}^{2}\xi P_{R})),\label{XXX}
\end{align}
for some $0\leq\xi\leq1$. Then, (\ref{R2stepend}) becomes
\begin{align}
nR_{2} &  \leq I(X_{2}^{n};Y_{2}^{n})+\frac{\eta_{BC}n}{2}\log(2\pi
e(1+c_{2}^{2}P_{R}))-\frac{\eta_{BC}n}{2}\log(2\pi e(1+c_{2}^{2}\xi
P_{R}))+n\epsilon_{n}\\
&  =I(X_{2}^{n};Y_{2}^{n})+\eta_{BC}n\mathcal{C}\left(  \frac{c_{2}%
^{2}\overline{\xi}P_{R}}{1+c_{2}^{2}\xi P_{R}}\right)  +n\epsilon_{n}%
\end{align}
where we have used the fact that Gaussian distribution maximizes the entropy
term for a given variance constraint.

Now, consider (\ref{Fanonend}) for the bound on $R_{1}$ given in (\ref{R1out3})
which follows as,
\begin{align}
%\label{scaling}
nR_{1}  &  \leq I(W_{1};Y_{1}^{n},Y_{R1}^{\eta_{BC}n}%
|W_{2})+n\epsilon_{n}\label{cond_dec_b}\\\label{markovb}
&  \leq I(X_{1}^{n};Y_{1}^{n}|X_{2}^{n})+I(W_{1};Y_{R1}^{\eta_{BC}n}|Y_{1}%
^{n},W_{2})+n\epsilon_{n}\\\label{scaling}
&  =I(X_{1}^{n};Y_{1}^{n}|X_{2}^{n})+I(W_{1};\frac{c_{2}}{c_{1}}Y_{R1}%
^{\eta_{BC}n}|Y_{1}^{n},W_{2})+n\epsilon_{n}%
\end{align}
where (\ref{cond_dec_b}) is due to conditioning decreases entropy and
independence of $W_{1}$ and $W_{2}$, (\ref{markovb}) is from Markovity
$W_{1}\rightarrow X_{1}^{n}\rightarrow Y_{1}^{n}$ and $X_{2}^{n}$ is a
function of $W_{2}$, (\ref{scaling}) is since scaling does not change the
mutual information.

Since the capacity region of BC depends on the conditional marginal
distributions and noting that $c_{1}\geq c_{2}$, we can write $Y_{R2}%
^{\eta_{BC}n}=\frac{c_{2}}{c_{1}}Y_{R1}^{\eta_{BC}n}+\widehat{Z}_{R}%
^{\eta_{BC}n}$ where $\widehat{Z}_{R}^{\eta_{BC}n}$ is an iid. Gaussian noise
with variance $1-\frac{c_{2}^{2}}{c_{1}^{2}}$. From the conditional Entropy
Power Inequality, we now have
\begin{equation}
\label{epiYR2}2^{\frac{2}{\eta_{BC}n}h(Y_{R2}^{\eta_{BC}n}|Y_{1}^{n},W_{2}%
)}\geq2^{\frac{2}{\eta_{BC}n}h\left(  \frac{c_{2}}{c_{1}}Y_{R1}^{\eta_{BC}%
n}|Y_{1}^{n},W_{2}\right)  }+2^{\frac{2}{\eta_{BC}n}h\left(  \widehat{Z}%
_{R}^{\eta_{BC}n}|Y_{1}^{n},W_{2}\right)  }.
\end{equation}
Also, for the condition $a_{12}\leq1$, we have,
\begin{align}
\label{hYR2ain}h(Y_{R2}^{\eta_{BC}n}|Y_{1}^{n},W_{2})  &  =h(Y_{R2}^{\eta_{BC}%
n}|X_{1}^{n}+Z_{1}^{n},W_{2})\\\label{hYR2a}
&  =h(Y_{R2}^{\eta_{BC}n}|X_{1}^{n}+Z_{2}^{n},W_{2})\\\label{hYR2b}
&  \leq h(Y_{R2}^{\eta_{BC}n}|a_{12}X_{1}^{n}+Z_{2}^{n},W_{2})\\\label{hYR2c}
&  = h(Y_{R2}^{\eta_{BC}n}|Y_{2}^{n},W_{2})\\\label{hYR2d}
&  = \frac{\eta_{BC}n}{2}\log(2\pi e(1+c_{2}^{2}\xi P_{R}))
\end{align}
where (\ref{hYR2ain}) is from the fact that $a_{21}X_{2}^{n}$ is a function of
$W_{2}$, (\ref{hYR2a}) follows from the independence of $Z_{R1}^{\eta_{BC}n}$,
$Z_{1}^{n}$, and $Z_{2}^{n}$, (\ref{hYR2b}) is due to the Markov chain,
$a_{12}X_{1}^{n}+Z_{2}^{n}\rightarrow X_{1}^{n}+Z_{2}^{n},W_{2}\rightarrow
Y_{R2}^{\eta_{BC}n}$ for the fact that $a_{12}\leq1$, (\ref{hYR2c}) is true
since $X_{2}^{n}$ is a function of $W_{2}$, and (\ref{hYR2d}) from
(\ref{XXX}). Then, using (\ref{epiYR2}), (\ref{hYR2d}), and noticing that
$h\left(  \widehat{Z}_{R}^{\eta_{BC}n}|Y_{1}^{n},W_{2}\right)  =\frac
{\eta_{BC}n}{2}\log\left(  2\pi e(1-\frac{c_{2}^{2}}{c_{1}^{2}})\right)  $, we
obtain,
\begin{align}
\label{h(cYR1)}h\left(  \frac{c_{2}}{c_{1}}Y_{R1}^{\eta_{BC}n}|Y_{1}^{n}%
,W_{2}\right)   &  \leq\frac{\eta_{BC}n}{2}\log\left(  2\pi e\left(  c_{2}%
^{2}\xi P_{R} + \frac{c_{2}^{2}}{c_{1}^{2}}\right)  \right)  .
\end{align}
So that, recalling (\ref{scaling}) and considering the inequality
(\ref{h(cYR1)}), we get,
\begin{align}
nR_{1}  &  \leq I(X_{1}^{n};Y_{1}^{n}|X_{2}^{n})+h(\frac{c_{2}}{c_{1}}%
Y_{R1}^{\eta_{BC}n}|Y_{1}^{n},W_{2})-h(\frac{c_{2}}{c_{1}}Z_{R1}^{\eta_{BC}%
n})+n\epsilon_{n}\\
&  \leq I(X_{1}^{n};Y_{1}^{n}|X_{2}^{n})+\eta_{BC}n\mathcal{C}\left(
c_{1}^{2}\xi P_{R}\right)  +n\epsilon_{n},
\end{align}
which recovers (\ref{R1out3}) and completes the proof.

Finally, we obtain the bound (\ref{R1_p8}) by continuing from (\ref{markovb})
\begin{align}
nR_{1} &  \leq H(W_{1}|W_{2})\label{markov-p8}\\
&  \leq I(W_{1};Y_{1}^{n},Y_{R1}^{\eta_{BC}n}|W_{2})+n\epsilon_{n}\\
&  \leq I(X_{1}^{n};Y_{1}^{n}|X_{2}^{n})+I(W_{1};Y_{R1}^{\eta_{BC}n}|Y_{1}%
^{n},W_{2})+n\epsilon_{n}\\
&  \leq I(X_{1}^{n};Y_{1}^{n}|X_{2}^{n})+h(Y_{R1}^{\eta_{BC}n})-h(Z_{R1}%
^{\eta_{BC}n})+n\epsilon_{n}\label{cond_dec_p8}\\
&  \leq I(X_{1}^{n};Y_{1}^{n}|X_{2}^{n})+n\eta_{BC}\mathcal{C}(c_{1}^{2}%
P_{R})+n\epsilon_{n}.\label{gaussp8}%
\end{align}
where (\ref{cond_dec_p8}) is true since conditioning decreases entropy and
(\ref{gaussp8}) is from the fact that Gaussian distribution maximizes the
entropy for given variance constraints. Using similar steps, we obtain the
rate for $R_{2}$ in (\ref{R2_p8}).$\Box$

\section*{G. Proof of Proposition 8}

The converse follows from (\ref{R1out2}), (\ref{R1out3}), and
(\ref{R2outc_end_b}) as
\begin{align}
R_{1}+R_{2} &  \leq\frac{1}{n}I(X_{1}^{n};Y_{1}^{n}|X_{2}^{n})+\frac{1}%
{n}I(X_{2}^{n};Y_{2}^{n})+\min\{\eta_{MAC}\mathcal{C}(b_{1}^{2}P_{1R}%
),\eta_{BC}\mathcal{C}(c_{1}^{2}\xi P_{R})\}\nonumber\label{epi_b}\\
&  +\eta_{BC}\mathcal{C}\left(  \frac{c_{2}^{2}\overline{\xi}P_{R}}%
{1+c_{2}^{2}\xi P_{R}}\right)  \\
&  \leq\mathcal{C}(a_{12}^{2}P_{1}+P_{2})+\mathcal{C}(P_{1})-\mathcal{C}%
(a_{12}^{2}P_{1})\nonumber\\
&  +\min\big\{\eta_{MAC}\mathcal{C}(b_{1}^{2}P_{1R}),\eta_{BC}\mathcal{C}%
(c_{1}^{2}\xi P_{R})\big\}+\eta_{BC}\mathcal{C}\left(  \frac{c_{2}%
^{2}\overline{\xi}P_{R}}{1+c_{2}^{2}\xi P_{R}}\right)  \\
&  =\mathcal{C}(P_{1})+\mathcal{C}\left(  \frac{P_{2}}{1+a_{12}^{2}P_{1}%
}\right)  +\min\big\{\eta_{MAC}\mathcal{C}(b_{1}^{2}P_{1R}),\eta
_{BC}\mathcal{C}(c_{1}^{2}\xi P_{R})\big\}\nonumber\\
&  +\eta_{BC}\mathcal{C}\left(  \frac{c_{2}^{2}\overline{\xi}P_{R}}%
{1+c_{2}^{2}\xi P_{R}}\right)  \label{TypeIIoutbnd}%
\end{align}
where (\ref{epi_b}) is from the worst-case noise result of \cite{worst-case}
applied for $a_{12}\leq1$. For $\eta_{MAC}\mathcal{C}(b_{1}^{2}P_{1R})\geq
\eta_{BC}\mathcal{C}(c_{1}^{2}P_{R})$, (\ref{TypeIIoutbnd}) is maximized for
$\xi=1$ since $c_{1}\geq c_{2}$, and hence the outer bound becomes
\begin{align}\label{Routprop11}
R_{1}+R_{2}\leq\mathcal{C}(P_{1})+\mathcal{C}\left(  \frac{P_{2}}{1+a_{12}%
^{2}P_{1}}\right)  +\eta_{BC}\mathcal{C}(c_{1}^{2}P_{R}).
\end{align}
\bigskip

For achievability, we use a special case of the coding scheme (\ref{codewords}%
) in which, over the IC, $S_{1}$ transmits private information only via a
Gaussian codebook $X_{1}^{n}(W_{1})=X_{1p}^{n}(W_{1p})$, and $S_{2}$ transmits
common information using a Gaussian codebook $X_{2}^{n}(W_{2})=X_{2c}%
^{n}(W_{2c^{\prime\prime}})$. Only private messages ($W_{1R},W_{2R}$) are sent
over the OBRC by using standard Gaussian codebooks and MAC decoding at the
relay, and superposition coding at the relay. The proof then follows similarly
to Proposition 2, Appendix B, by accounting for the capacity regions of MAC
and BC Gaussian channels (see, e.g., \cite{Kramer_booklet}). Specifically, we
obtain the following rates,
\begin{align}
R_{1p} &  \leq\mathcal{C}(P_{1})\label{ach_fin_a}\\
R_{2c} &  \leq\mathcal{C}\left(  \frac{P_{2}}{1+a_{12}^{2}P_{1}}\right)  \\
R_{1p}+R_{2c} &  \leq\mathcal{C}(P_{1}+a_{21}^{2}P_{2})\\
R_{1R} &  \leq\min\left\{  \eta_{MAC}\mathcal{C}(b_{1}^{2}P_{1R}),\eta
_{BC}\mathcal{C}\left(  c_{1}^{2}\xi P_{1R}\right)  \right\}  \\
R_{2R} &  \leq\min\left\{  \eta_{MAC}\mathcal{C}(b_{2}^{2}P_{2R}),\eta
_{BC}\mathcal{C}\left(  \frac{c_{2}^{2}\overline{\xi}P_{1R}}{1+c_{2}^{2}\xi
P_{R}}\right)  \right\}  \\
R_{1R}+R_{2R} &  \leq\eta_{MAC}\mathcal{C}\left(  b_{1}^{2}P_{1R}+b_{2}%
^{2}P_{2R}\right)  \label{ach_fin_end}%
\end{align}
Then, setting $R_{1}=R_{1p}+R_{1R}$, $R_{2}=R_{2c}+R_{2R}$ with $\xi=1$,
$\eta_{MAC}\mathcal{C}(b_{1}^{2}P_{1R})\geq\eta_{BC}\mathcal{C}(c_{1}^{2}%
P_{R})$ and $a_{21}\geq\sqrt{\frac{1+P_{1}}{1+a_{12}^{2}P_{1}}}$, the outer
bound (\ref{Routprop11}) is achievable, hence we obtain the sum capacity.

\section*{H. Proof of Proposition 9}

The proof follows that in Appendix G. For $\eta_{MAC}\mathcal{C}(b_{1}%
^{2}P_{1R})<\eta_{BC}\mathcal{C}(c_{1}^{2}P_{R})$, denote $\xi^{\ast}$ as the
optimal parameter that maximizes (\ref{TypeIIoutbnd}) so that we get
\begin{align}
R_{1}+R_{2} &  \leq\mathcal{C}(P_{1})+\mathcal{C}\left(  \frac{P_{2}}%
{1+a_{12}^{2}P_{1}}\right)  +\min\big\{\eta_{MAC}\mathcal{C}(b_{1}^{2}%
P_{1R}),\eta_{BC}\mathcal{C}(c_{1}^{2}\xi^{\ast}P_{R}%
)\big\}\nonumber\label{Out_bnd_df}\\
&  +\eta_{BC}\mathcal{C}\left(  \frac{c_{2}^{2}\overline{\xi}^{\ast}P_{R}%
}{1+c_{2}^{2}\xi^{\ast}P_{R}}\right)
\end{align}
where $\xi^{\ast}+\overline{\xi}^{\ast}=1$. Moreover, from (\ref{ach_fin_a}%
)-(\ref{ach_fin_end}), for the conditions $a_{21}\geq\sqrt{\frac{1+P_{1}%
}{1+a_{12}^{2}P_{1}}}$, $\eta_{MAC}\mathcal{C}(b_{1}^{2}P_{1R})<\eta
_{BC}\mathcal{C}(c_{1}^{2}P_{R})$ and $\eta_{MAC}\mathcal{C}\left(
\frac{b_{2}^{2}P_{2R}}{1+b_{1}^{2}P_{1R}}\right)  \geq\eta_{BC}C\left(
\frac{c_{2}^{2}\overline{\xi}^{\ast}P_{R}}{1+c_{2}^{2}\xi^{\ast}P_{R}}\right)
$ and with the power split of $\xi^{\ast}P_{R}$ allocated at the relay for the
transmission of $W_{1R}$ (and $\overline{\xi}^{\ast}P_{R}$ for $W_{2R}$), the
achievable sum-rate obtained by $R_{1}=R_{1p}+R_{1R}$, $R_{2}=R_{2c}+R_{2R}$,
is equal to the outer bound (\ref{Out_bnd_df}).

\section{I. Proof of Proposition 10}

The achievable region is obtained similarly to Appendices B and G. Source
$S_{1}$ transmits private message only, i.e., $X_{1}^{n}(W_{1})=X_{1p}%
^{n}(W_{1p})$ over the IC and independent private message over the OBRC
$W_{1R}$ via Gaussian codebooks. Source $S_{2}$ transmits common messages
$(W_{2c^{\prime}},W_{2c^{\prime\prime}})$ over the IC ($X_{2}^{n}%
(W_{2})=X_{2c}^{n}(W_{2c^{\prime}},W_{2c^{\prime\prime}})$), and the private
message $W_{2R}$ is transmitted via the OBR along with $W_{2c^{\prime}}$
(interference forwarding). Then, the following conditions are easily seen to
provide an achievable region
\begin{align}
R_{1p} &  \leq\mathcal{C}(P_{1})\\
R_{2c^{\prime\prime}}+R_{1p} &  \leq\mathcal{C}(P_{1}+a_{21}^{2}P_{2})\\
R_{2c} &  \leq\mathcal{C}\left(  \frac{P_{2}}{1+a_{12}^{2}P_{1}}\right)  \\
R_{1R} &  \leq\eta_{MAC}\mathcal{C}(b_{1}^{2}P_{1R})\\
R_{2c^{\prime}}+R_{2R} &  \leq\eta_{MAC}\mathcal{C}(b_{2}^{2}P_{2R})\\
R_{1R}+R_{2c^{\prime}}+R_{2R} &  \leq\eta_{MAC}\mathcal{C}(b_{1}^{2}%
P_{1R}+b_{2}^{2}P_{2R})\\
R_{2c^{\prime}}+R_{1R} &  \leq\eta_{BC}\mathcal{C}(c_{1}^{2}\xi P_{R})\\
R_{2R} &  \leq\eta_{BC}\mathcal{C}\left(  \frac{c_{2}^{2}\overline{\xi}P_{R}%
}{1+c_{2}^{2}\xi P_{R}}\right)
\end{align}
Using Fourier-Motzkin elimination method, with the fact that $R_{1}%
=R_{1p}+R_{1R}$, $R_{2c}=R_{2c^{\prime}}+R_{2c^{\prime\prime}}$, and
$R_{2}=R_{2c}+R_{2R}$, the achievable region in Proposition 10 can be obtained.
Now, for $b_{2},c_{1}\rightarrow\infty$, the achievable region becomes
\begin{align}
%\label{prop14R2}
R_{1} &  \leq\mathcal{C}(P_{1})+\eta_{MAC}\mathcal{C}%
(b_{1}^{2}P_{1R})\label{prop14R1}\\
R_{2} &  \leq\mathcal{C}\left(  \frac{P_{2}}{1+a_{12}^{2}P_{1}}\right)
+\eta_{BC}\mathcal{C}\left(  c_{2}^{2}P_{R}\right)\label{prop14R2}
\end{align}
since the overall region is maximized for $\xi=0$ for $b_{2},c_{1}%
\rightarrow\infty$.

For the outer bound, we use the bound on IC-OBR Type-II given in
(\ref{TypeIIoutbnd}). Again, as $b_{2},c_{1}\rightarrow\infty$, the outer
bound is maximized for $\xi=0$, and the achievable sum-rate $R_{1}+R_{2}$
obtained by the summation of (\ref{prop14R1}) and (\ref{prop14R2}) is equal to
the outer bound (\ref{TypeIIoutbnd}), thus concluding the proof.$\Box$

\section{J. Proof of Proposition 11}
\begin{figure}[t]
\begin{center}
\includegraphics[width=10 cm]{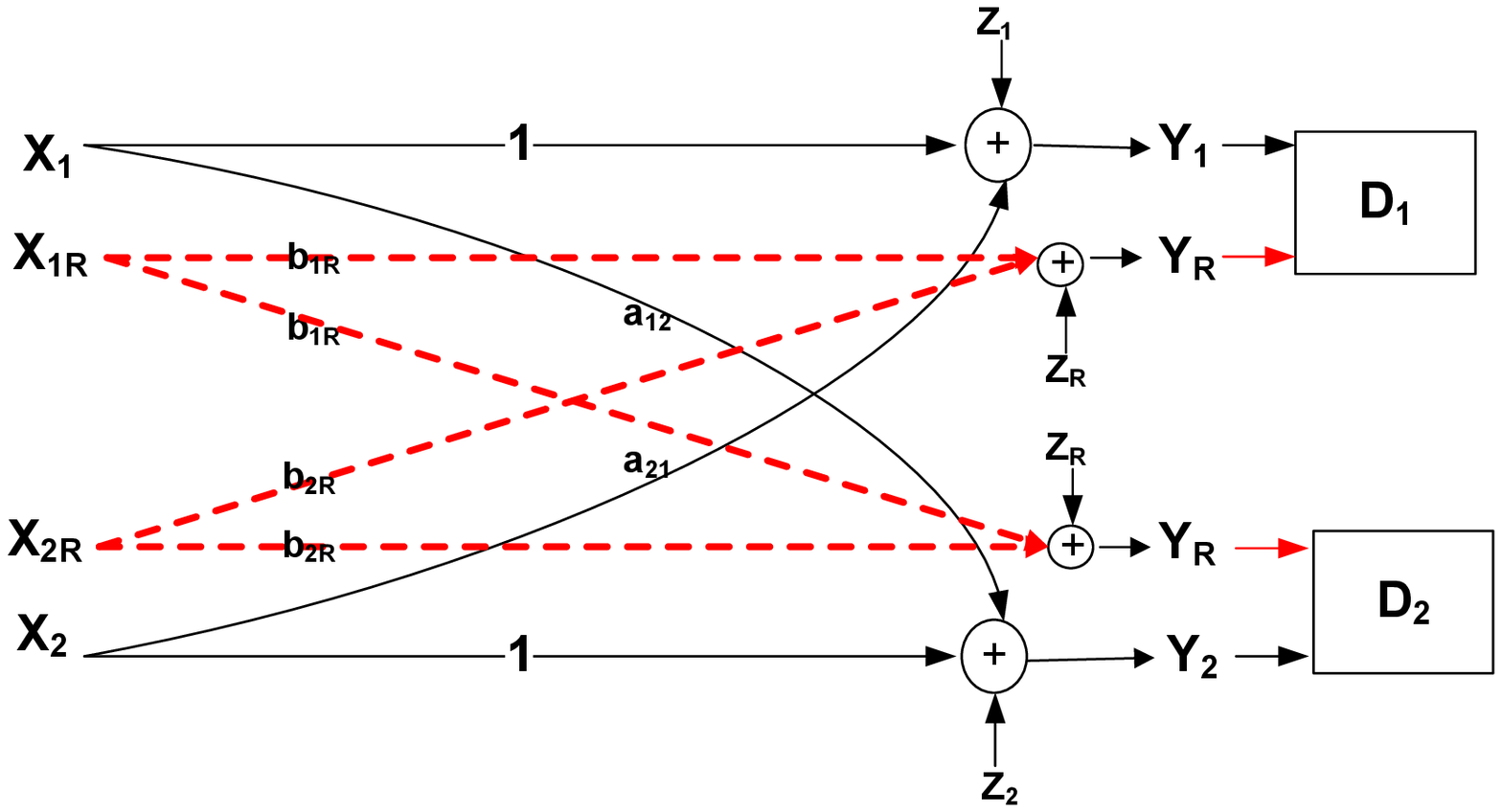}
\end{center}
\caption{Equivalent model for IC-OBR Type II channel for $c_{1},c_{2}%
\rightarrow\infty$.}%
\label{Fig:Eq_mod_CF}%
\end{figure}

The achievable rates follow from standard arguments assuming Wyner-Ziv
compression at the relay with Gaussian test channels (see, e.g.,
\cite{Kramer_booklet}). The converse for $a_{12}\geq1,$ $a_{21}\geq1$ follows
from the results in \cite{MIMO_IC}. Specifically, assume $c_{1},c_{2}%
\rightarrow\infty$, so that we obtain an equivalent model as shown in Fig.
\ref{Fig:Eq_mod_CF}. The model in Fig. \ref{Fig:Eq_mod_CF} is in fact
equivalent to a $2\times2$ MIMO interference channel whose channel matrices,
following the notation in \cite{MIMO_IC} are given by $\mathbf{H_{1}}=\left[
\begin{matrix}
1 & 0\\
0 & b_{1}%
\end{matrix}
\right]  $, $\mathbf{H_{2}}=\left[
\begin{matrix}
a_{21} & 0\\
0 & b_{2}%
\end{matrix}
\right]  $, $\mathbf{H_{3}}=\left[
\begin{matrix}
a_{12} & 0\\
0 & b_{1}%
\end{matrix}
\right]  $, and $\mathbf{H_{4}}=\left[
\begin{matrix}
1 & 0\\
0 & b_{2}%
\end{matrix}
\right]  .$ Notice that such equivalence is due to the fact that noise
correlations are immaterial in terms of the capacity region. For this channel,
the assumed conditions $a_{12}\geq1$ and $a_{21}\geq1$ imply the strong
interference regime $\mathbf{H}_{2}^{\dag}\mathbf{H}_{2}\succeq\mathbf{H}%
_{4}^{\dag}\mathbf{H}_{4}$ and $\mathbf{H}_{3}^{\dag}\mathbf{H}_{3}%
\succeq\mathbf{H}_{1}^{\dag}\mathbf{H}_{1}$, so that the capacity region can
be found from \cite{MIMO_IC} as
\begin{subequations}
\label{CF_cap_regend_a}%
\begin{align}
R_{1} &  \leq\mathcal{C}(P_{1})+\eta_{MAC}\mathcal{C}(b_{1}^{2}P_{1R}%
)\label{rtotdfob1}\\
R_{2} &  \leq\mathcal{C}(P_{2})+\eta_{MAC}\mathcal{C}(b_{2}^{2}P_{2R})\\
R_{1}+R_{2} &  \leq\mathcal{C}(P_{1}+a_{21}^{2}P_{2})+\eta_{MAC}%
\mathcal{C}(b_{1}^{2}P_{1R}+b_{2}^{2}P_{2R})\\
R_{1}+R_{2} &  \leq\mathcal{C}(a_{12}^{2}P_{1}+P_{2})+\eta_{MAC}%
\mathcal{C}(b_{1}^{2}P_{1R}+b_{2}^{2}P_{2R}).\label{CF_cap_regend}%
\end{align}
This proves the desired result.
\end{subequations}

\end{document}